\documentclass[12pt,manuscript]{aastex}
\usepackage{lineno}
\usepackage{longtable}
\linenumbers*[1]

\newcommand{\mearth}{M$_{\oplus}$}
\newcommand{\rearth}{R$_{\oplus}$}

\newcommand{\rsun}{R$_{\odot}$}
\newcommand{\rjupiter}{R$_{J}$}
\newcommand{\etaearth}{$\eta_{\oplus}$}

\slugcomment{Accepted to \emph{The Astrophysical Journal} on 26 April 2013}

\shorttitle{Candidate Habitable Planets around \emph{Kepler} Stars}
\shortauthors{Gaidos}

\begin{document}


\title{Candidate Planets in the Habitable Zones of \emph{Kepler} Stars}

\author{Eric Gaidos} 
\affil{Department of Geology and Geophysics, University of Hawai`i at M\={a}noa, Honolulu, HI 96822}

\email{gaidos@hawaii.edu}

\begin{abstract}
  A key goal of the \emph{Kepler} mission is the discovery of
  Earth-size transiting planets in ``habitable zones'' where stellar
  irradiance maintains a temperate climate on an Earth-like planet.
  Robust estimates of planet radius and irradiance require accurate
  stellar parameters, but most \emph{Kepler} systems are faint, making
  spectroscopy difficult and prioritization of targets desirable.  The
  parameters of 2035 host stars were estimated by Bayesian analysis
  and the probabilities $p_{HZ}$ that 2738 candidate or confirmed
  planets orbit in the habitable zone were calculated.  Dartmouth
  Stellar Evolution Program models were compared to photometry from
  the \emph{Kepler} Input Catalog, priors for stellar mass, age,
  metallicity and distance, and planet transit duration.  The analysis
  yielded probability density functions for calculating confidence
  intervals of planet radius and stellar irradiance, as well as
  $p_{HZ}$.  Sixty-two planets have $p_{HZ} > 0.5$ and a most probable
  stellar irradiance within habitable zone limits.  Fourteen of these
  have radii less than twice the Earth; the objects most resembling
  Earth in terms of radius and irradiance are KOIs 2626.01 and
  3010.01, which orbit late K/M-type dwarf stars.  The fraction of
  \emph{Kepler} dwarf stars with Earth-size planets in the habitable
  zone (\etaearth{}) is 0.46, with a 95\% confidence interval of
  0.31-0.64.  Parallaxes from the \emph{Gaia} mission will reduce
  uncertainties by more than a factor of five and permit definitive
  assignments of transiting planets to the habitable zones of
  \emph{Kepler} stars.
\end{abstract}

\section{Introduction}

The \emph{Kepler} mission was launched in March 2009 with a mission to
find Earth-size planets in the circumstellar ``habitable zone'' (HZ)
of solar-type stars \citep{Borucki2010Sci}.  Broadly speaking, the HZ is
considered the range of orbital semimajor axes over which the surface
temperature on an Earth-like planet would permit liquid water.  A
narrower definition, adopted here, is that it is the range of stellar
irradiance between the runaway ``wet'' greenhouse limit - beyond which
a water-vapor saturated N$_2$-CO$_2$ atmosphere cannot radiate, and
the CO$_2$ ``snowball'' limit below which this greenhouse gas
condenses from an Earth-like atmosphere onto the poles
\citep{Kasting1993,Ishiwatari2007}.  This definition makes assumptions
about planetary albedo, rotation rate \citep{Spiegel2008}, orbital
eccentricity and obliquity \citep{Williams2003}, extent of oceans
\citep{Abe2011}, and thickness and composition of the atmosphere
\citep{Pierrehumbert2011ApJ734}.  Many other factors besides stellar
irradiation determine habitability \citep{Gaidos2005}.  A planet in
the canonical HZ may not be Earth-like, e.g., if it is geologically
inactive \citep{Kite2009}, and there may be habitable environments
outside the HZ, e.g. in the interiors of icy satellites
\citep{Reynolds1987}.  Nevertheless, an orbit in this HZ is a useful
criterion for selecting objects for follow-up observations. Such
prioritization is essential given that there are thousands of faint
($\sim$15th magnitude) \emph{Kepler} systems that would require
impractical amounts of telescope time to study.

\citet{Borucki2011ApJ736} published a catalog of 54 (out of 1235) candidate
planets or \emph{Kepler} Objects of Interest (KOIs) with equilibrium
emitting temperatures between 273 and 373~K, assuming an Earth-like
albedo of 0.3.  \citet{Kaltenegger2011} noted the importance of
albedo, specifically cloud cover, to equilibrium temperature, and
computed inner and outer HZ boundaries based on the stellar
irradiation criteria derived by \citet{Selsis2007a} for high H$_2$O
and high CO$_2$ atmospheres, respectively.  They identified 76
possible habitable planets, depending on the assumed fractional cloud
cover.  They found that many of the \citet{Borucki2011ApJ736} candidates
were too hot for this habitability criterion and pointed out that
errors in stellar parameters contribute most to the uncertainty of
whether a planet orbits within the HZ.

Subsequently, a larger catalog (2300 KOIs, including some that are
confirmed planets), was released \citep{Batalha2013}.  Stellar
parameters for KOI hosts, i.e. mass $M_*$ and radius $R_*$, were
determined by fitting Yale-Yonsei model isochrones
\citep{Demarque2004} to values of effective temperature ($T_*$),
surface gravity ($\log g$), and metallicity ([Fe/H]).  Stellar
parameters were derived from the photometry of the \emph{Kepler} Input
Catalog (KIC) and a model of stellar populations and Galactic
structure \citep{Brown2011AJ142}.  The \citet{Batalha2013} estimates of
mass and radii assumed gaussian-distributed errors and employed
standard deviations derived from a comparison between KIC-derived
parameters and spectroscopic values.  They revised the
\citet{Brown2011AJ142} estimates of $\log g$ and $R_*$ for many stars.
\citet{Batalha2013} assumed an albedo of 0.3 and efficient
redistribution of heat over a planet's surface, and identified 46
candidates with 185~K $< T_{eq} <$ 303~K.

However, the \citet{Brown2011AJ142} stellar parameters themselves are
uncertain and in some aspects problematic.  The vast majority of
\emph{Kepler} stars do not yet have measured parallaxes.  KIC
photometry must be corrected to place it in the Sloan system, and
KIC-based effective temperatures are about 200~K hotter than estimates
based on the infrared flux method \citep{Pinsonneault2012}.  Moreover,
uncertainties in stellar parameters, and hence incident irradiance,
can be markedly non-gaussian.  This is particularly true for
solar-type stars for which photometry is unable to distinguish between
main sequence and evolved (subgiant) stars
\citep{Brown2011AJ142,Gaidos2013}.  In such cases, standard deviations
have limited utility in assessing statistical confidence.

A more rigorous approach is to estimate a \emph{probability} that a
planet orbits in the HZ, i.e. that the irradiance falls between the
wet runaway greenhouse and CO$_2$ condensation limits.  This can be
done using the probability distribution function (PDF) of irradiance
calculated from PDFs of the stellar parameters.  The latter can be
generated by comparing stellar models to observational constraints
(i.e. photometry), calculating probabilities that the models can
explain the data, and conditioning these by Bayesian priors.  Each
model and its associated value for irradiance is assigned a posterior
probability, and the probability that the planet orbits in the HZ is
the sum of the probabilities for those models having irradiances
within the HZ limits, divided by the total probability for all models.

Bayesian estimation of stellar parameters has been applied to the KIC
\citep{Brown2011AJ142} as well as \emph{Hipparcos} stars
\citep{Bailer-Jones2011}\footnote{Alternative approaches to the use of
  broad-band photometry to derive stellar parameters are described in
  \citet{Ammons2006} and \citet{Belikov2008}.}.  The analysis
described here is distinguished by the use of corrected KIC
photometry, synthetic isochrones and photometry from the Dartmouth
Stellar Evolution database \citep{Dotter2008} (see also
\citet{Dressing2013}, and new priors that describe distributions with
mass (IMF), metallicity, age, and distance using recent models of the
Galaxy \citep{Vanhollebeke2009}.  In addition, it uses the duration
and probability of planet transits to constrain stellar density
\citep{Plavchan2012}.

I applied this procedure to the catalog of 2740 confirmed and
candidate planets around 2036 \emph{Kepler} stars released on 7
January 2013.  I estimated the expected fraction of stars with planets
orbiting in the HZ, and identified (candidate) planets with a
better-than-even chance of having such orbits.  I also cataloged
Earth- to Super Earth-size planets with lower but non-zero
probabilities.  These objects are high-priority targets for follow-up
observations to confirm the planets and better characterize their host
stars.

\section{Methods \label{sec.methods}}

\subsection{Algorithm \label{subsec.algorithm}}

I compared photometry for each star with sets of synthetic SDSS+2MASS
\emph{grizJHK$_s$} photometry from the isochrones of the Dartmouth
Stellar Evolution Program (DSEP) \citep{Dotter2008}.  With appropriate
choices of mixing length and initial helium and heavy element
fractions, DSEP is able to accurately reproduce the radius,
luminosity, and convective boundary of the Sun, as well as the radii
of fully convective stars in the hierarchical triple system KOI-126
\citep{Feiden2011}.  DSEP uses PHOENIX model stellar atmospheres as
boundary conditions; these LTE atmosphere models compare favorably to
non-LTE calculations and observations for stars cooler than 7000~K
\citep{Hauschildt1999}.

I compared up to six colors constructed with respect to the $r$
magnitude.  According to Bayes' theorem, the probability $P_i$ that
the $i$th model (hypothesis) is supported by the photometry is equal
to the probability that the colors ${c_j}$ can be produced by the
model, multipled by a prior function $p_i$.  Assuming
gaussian-distributed errors in photometry, that probability is
\begin{equation}
\label{eqn.prob}
P_i = p_i {\rm exp}\left[-\displaystyle\sum_j \frac{\left(c_j - k_jE^i_{B-V}-\hat{c_j}\right)^2}{2\sigma_j^2}\right],
\end{equation}
where the summation is over up to six colors, ${\hat{c_j}}$ are the
synthetic colors, ${\sigma_j}$ are the photometric errors, $k_j$ is
the interstellar reddening coefficient for the color, and $E^i_{B-V}$
is the amount of reddening that is assigned to a particular model and
star (see below).  The normalization in Eqn. \ref{eqn.prob} is
unimportant as it independent of the models and I identified the model
which has the largest value of $P$.  The prior is the product of
individual priors for the mass, age, distance, and metallicity of the
model, the intervening extinction, and, since at least one planet has
been detected around each of these stars, a constraint on stellar
density imposed by the duration of the transit \citep{Plavchan2012}.

Photometry and other data for the host stars of the KOIs were
extracted from the KIC catalog available at the MAST database.  KIC
\emph{griz} magnitudes were transformed to the Sloan system using the
corrections determined by \citet{Pinsonneault2012}.  Standard errors
for each bandpass were estimated using the expression $\sigma =
\sigma_0 10^{(m-m_0)/2.5}$, where
$\sigma_0=0.02,0.02,0.015,0.015,0.02,0.025,0.02$, and $m_0 =
15,15,15.3,15.3,13,11.75,10.8$, for \emph{grizJHK}, respectively
\citep{Brown2011AJ142,Cutri2003}.  For \emph{griz} the magnitude $m$ is
the \emph{Kepler} magnitude $K_p$ and for \emph{JHK$_s$} it is the
respective 2MASS magnitudes.  Errors in color were calculated by
assuming that errors in individual bandpasses are uncorrelated and
adding the two corresponding such errors in quadrature.

Minimization of $P$ with respect to $E_{B-V}$ leads to a formula for
the best-fit reddening for each model:
\begin{equation}
\label{eqn.reddening}
  E_{B-V} = \frac{\displaystyle\sum_jk_i\left(c_j - \hat{c}_j\right)/\sigma_j^2}{\displaystyle\sum_j k_j^2/\sigma_j^2}
\end{equation}
I adopted extinction coefficients $A$ of 3.758 ($g$), 2.565 ($r$),
1.874 ($i$), 1.377 ($z$), 0.272 ($J$), 0.173 ($H$), and ($K_s$), based
on \citet{Girardi2005} and \citet{Chen2007}.

I used the DSEP interpolator tool to construct a grid of isochrones
with [Fe/H] $\in [-1.5,+0.5]$, at intervals of 0.1 dex, $\alpha/Fe$
$\in [-0.2,+0.4]$, at intervals of 0.2 dex, and ages $\in [1,12]$~Gyr
at intervals of 0.5~Gyr.  All models used a helium fraction $Y = 0.245
+ 1.5Z$, where $Z$ is the total heavy element abundance.  I further
restricted the selection to stars with initial masses between 0.1 and
2 solar masses, as late M and O, B, and early A-type stars are absent
from the \emph{Kepler} target list \citep{Batalha2010}.  This
restriction reduced the total number of models considered to 657,347.

Once the best-fit model with the maximum $P$ was found, additional
models (typically a few dozen) with neighboring (difference less than
1.5 times the grid spacing) values of mass, age, [Fe/H], and
[$\alpha$/Fe] were identified.  A set of 100 linear interpolations
between the best-fit model and each of these neighboring models was
made and new probablities calculated using Eqn. \ref{eqn.prob}.  The
interpolation yielding the highest value of $P$ was recorded.

\subsection{Prior functions \label{sec.priors}}

Priors weight each DSEP model, i.e. each combination of initial mass,
metallicity, age, and distance (modulus).  The distance modulus for
each star/model combination was computed in the $r$-band, i.e. $\mu_r
= r - A_r E_{B-V} - \hat{M_r}$.  A uniform prior is adopted for for
the allowed range of [$\alpha$/Fe] between -0.2 and +0.4 dex.  As a
prior for initial stellar masses I adopt the tripartate initial mass
function (IMF) of \citet{Kroupa2002}.  Priors for age, metallicity,
and distance modulus $\mu = m - M$ were constructed using the
distributions of dwarf stars ($\log g > 4$) with $K_p < 16$
synthesized using TRILEGAL \citep{Vanhollebeke2009}.  TRILEGAL
accurately reproduces star counts over a wide range of magnitudes to
very low galactic latitudes \citep{Girardi2012}.  The simulated
population was restricted to dwarfs to reflect the criteria of the
selection of \emph{Kepler} targets \citep{Batalha2010}.  The (mostly
default) values for key TRILEGAL parameters are the same as used in
\citet{Gaidos2013}.

\begin{figure}
\epsscale{0.45}
\plotone{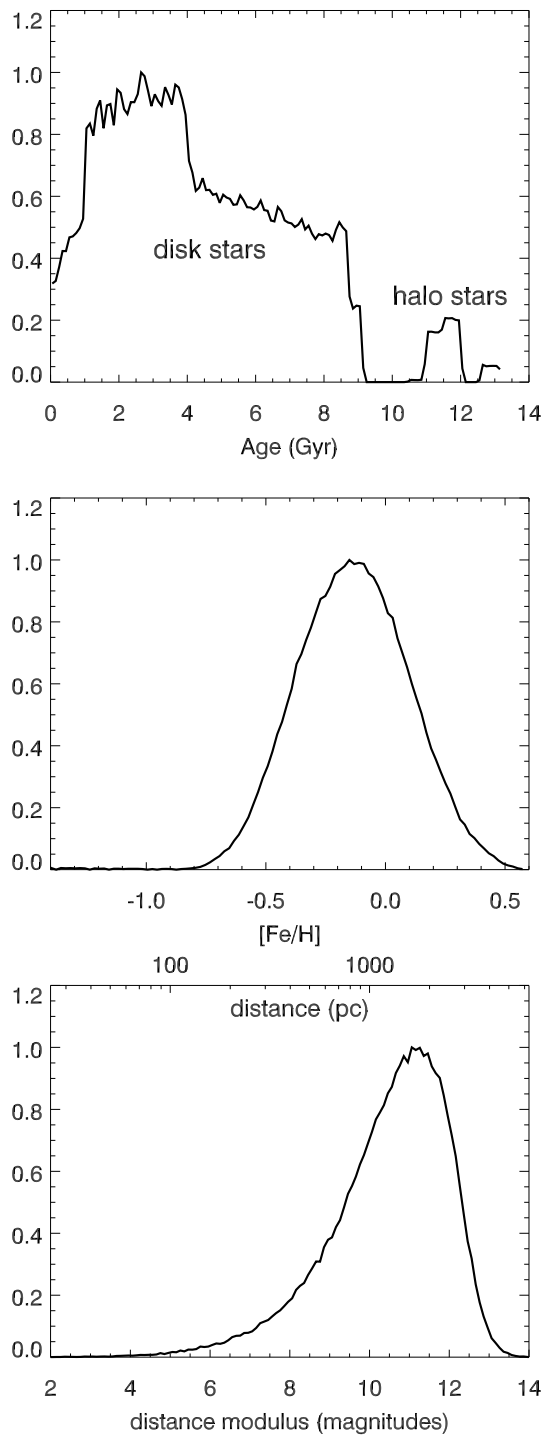}
\caption{Priors of age, metallicity, and distance modulus generated
  from 428,792 TRILEGAL-simulated stars with $K_p < 16$ and $\log g >
  4$ in the \emph{Kepler} field.  The distributions of all synthetic
  stars is presented here, but only the $\sim$10\% of stars with
  Galactic latitudes within 0.5~deg of a \emph{Kepler} star of
  interest are used to generate actual priors.  \label{fig.priors}}
\end{figure}

The resulting prior distributions (Fig. \ref{fig.priors}) have a
median metallicity of -0.13, median age of 3.9~Gyr, and median $\mu =
11.2$ ($\sim$1740~pc).  The age distribution is complex because the
population includes halo stars, which formed 11-12~Gyr ago, and disk
stars, which started forming 9~Gyr ago in these simulations.  TRILEGAL
models the star formation rate in the disk in two steps, with the
second occurring at about the epoch of the Sun's formation.  The
paucity of stars younger than 1~Gyr is partly due to the fact that the
\emph{Kepler} field probes the stellar population that is $>100$~pc
above the Galactic plane.  Of course, stellar ages, metallicities, and
distances are interrelated, but here they are used separately,
providing broad constraints on the possible ranges of stellar
parameters.  The distance distribution is particularly important in
allowing the finite scale height of the Galactic disk to prevent
Malmquist bias from selecting arbitrarily distant and luminous stars.

The \emph{Kepler} field is $\sim$10 degrees wide and close to the
Galactic plane ($b \sim 13$ deg), so the stellar populations that are
probed will vary significantly across the field.  Using TRILEGAL, I
synthesized the stellar population over a square degree centered at
each of 84 \emph{Kepler} half-CCD fields.  Only those synthetic
populations for CCD field centers with $b$ within 0.5 deg of a given
\emph{Kepler} star (about 10\% of the total) were used to calculate
priors for age, metallicity, and distance.

A prior for extinction $E_{B-V}$ necessarily involves information
about the distribution of both stars and dust along each line of
sight.  However, by assuming that the spatial distributions of stars
and dust are the same, the prior becomes particularly simple: a
uniform distribution between 0 and total ($\infty$) extinction along
the line of sight (see Appendix A).  I found improved agreement with
spectroscopy (Section \ref{subsec.compare}) by conditioning $E_{B-V}$
with a uniform prior between 0 and $E_{B-V}(\infty)[1-exp(-z/h)]$,
where $z$ is the vertical galactic distance above the Sun based on
$\mu_r$, and $h$ is the dust scale height
\citep[$\sim$200~pc;][]{Drimmel2001}.  I adopted the
\citet{Schlegel1998} Galactic reddening maps and interpolated the
total extinction at the coordinates of each star using the IDL tools
provided by the Princeton website.  Models with optimal $E_{B-V}$
values outside this range are allowed, but reddening is limited to the
maximum value and the models are penalized for the resulting
disagrement between measured and model colors (Eqn. \ref{eqn.prob}).

\subsection{Constraints from the planet transit \label{sec.transit}}

\begin{figure}
\epsscale{0.7}
\plotone{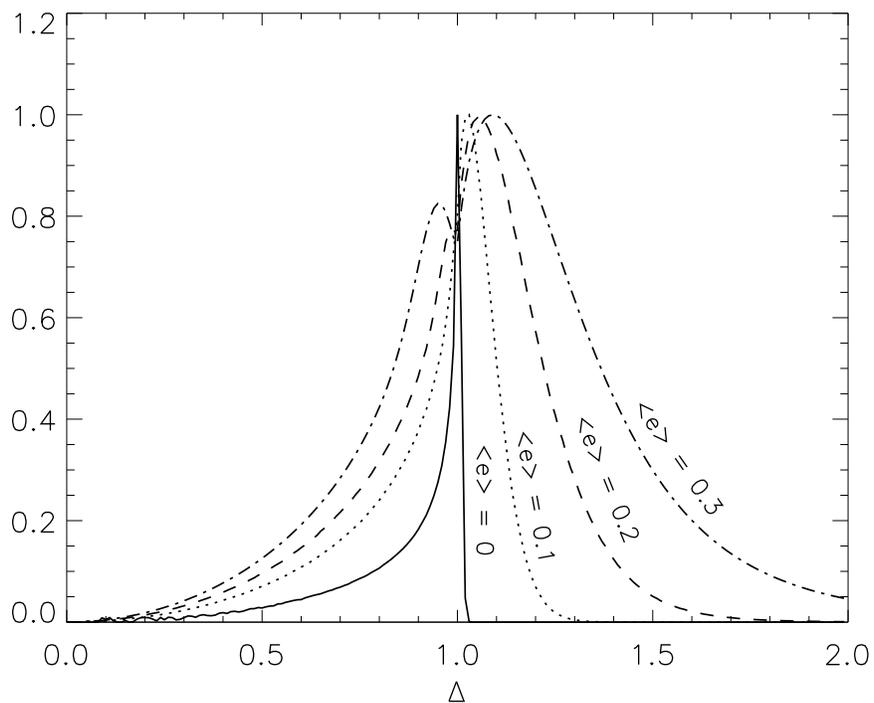}
\caption{Priors on the planet transit duration as a function of the
  parameter $\Delta \equiv D/(\tau^{2/3}P_K^{1/3})$, where $D$ is the
  transit duration, $\tau$ is the stellar free-fall time, and $P_K$ is
  the Keplerian orbital period, for circular orbits (solid line) and
  orbits with Rayleigh-distributed eccentricities with means of 0.1,
  0.2, and 0.3.  \label{fig.duration}}
\end{figure}

The transit duration $\tau$ and orbital period $P_K$ of a transiting
planet constrain stellar density \citep{Plavchan2012} and can be used
as an additional prior for stellar models.  In the case of
\emph{Kepler} low-cadence data, the constraint is weakened by a lack
of information about the orbit, specifically independent determination
of the orbital eccentricity $e$ and the transit impact parameter $b$.  The
transit duration $D$ is:
\begin{equation}
\label{eqn.duration}
D = \tau^{2/3} P_K^{1/3}\frac{\sqrt{(1-e^2)(1-b^2)}}{1 + e \cos \phi},
\end{equation}
where the stellar free-fall time is $\tau = 2\sqrt{\hat{R_*}^3/(\pi G
  \hat{M_*})}$, $G$ is the gravitational constant, and $\phi$ is the
argument of periastron relative to the line of sight to the star.
Given $D$ and $P_K$ and a value for $\tau$ for each stellar model, $e$
can be written as a function of $b$ and $\phi$:
\begin{equation}
\label{eqn.eccetricity}
e(b,\phi,\Delta) = \frac{\sqrt{\left(1-b^2\right)\left(1 -b^2 - \Delta^2\sin^2\phi\right)} - \Delta^2 \cos \phi}{1 - b^2 + \Delta^2 \cos^2 \phi},
\end{equation}
where $\Delta \equiv D/(\tau^{2/3}P^{1/3})$.  The eccentricity was
calculated over a uniform grid of $b \in [0,1]$ and $\phi \in
[0,2\pi]$.  Each value of $e$ was assigned a probability, i.e. values
not $\in [0,1]$ were assigned zero and others were assigned
probabilities from a prior distribution of $e$.  A Rayleigh
distribution,
\begin{equation}
\label{eqn.eccentricity}
n(e) = \frac{e}{\sigma^2}e^{-e^2/(2\sigma^2)},
\end{equation}
was assumed, with $\langle e \rangle = \sigma \sqrt{\pi/2}$.  Such a
distribution has been used in a previous analysis of \emph{Kepler}
transit durations \citep{Moorhead2011a} and is motivated by dynamical
theory \citep{Juric2008}.  Then the prior for the $i$th model from the
duration of the transit is
\begin{equation}
p_i = \frac{1}{2\pi} \int_0^1 db \int_0^{2\pi} d\phi \: n\left(e(b,\phi; \Delta_i)\right).
\end{equation}
To account for finite errors in transit duration, the prior can be
calculated using multiple Monte Carlo-generated values of $D$ is
repeated and then averaged.  In the case of a multi-planet system
$j=1...N$, the product of the individual transit duration priors
$\prod_j p_{ij}(\Delta_{ij})$ was used.  A value of $\sigma =
0.2\sqrt{2/\pi}$ for the dispersion in eccentricities was used based
on \citet{Moorhead2011a}.  Figure \ref{fig.duration} plots the prior
for four values of $\langle e \rangle$.

\subsection{Probability that a planet orbits in the habitable zone}

Orbit-averaged irradiation is only weakly dependent on eccentricity
for near-circular orbits.  I assume near-circular orbits in which case
the orbit-averaged irradiation in terrestrial units is approximately
\begin{equation}
I \approx \frac{\hat{L_*}}{L_{\odot}}\left(\frac{P_K}{365.24~{\rm d}}\right)^{4/3}\left(\frac{\hat{M_*}}{M_{\odot}}\right)^{2/3}.
\end{equation}
A planet is defined to be in the HZ if $I_{out} < \bar{I} < I_{in}$,
where the irradiance of the inner edge of the HZ for a 50\%
cloud-covered planet with efficient heat re-distribution is
\citep{Selsis2007a}
\begin{equation}
I_{in} = \left[0.68 - 2.7619 \times 10^{-5} \Theta - 3.8095 \times 10^{-9} \Theta^2\right]^{-2},
\end{equation}
and the outer edge is:
\begin{equation}
I_{out} = \left[1.95 - 1.3786 \times 10^{-4} \Theta - 1.4286 \times 10^{-9} \Theta^2\right]^{-2},
\end{equation}
where $\Theta \equiv \hat{T_*} - 5700$.  These functions account for
two major factors that introduce a dependence of the HZ boundaries on
the stellar spectrum (and hence effective temperature): the dependence
of Rayleigh scattering on wavelength, and the strong absorption by
H$_2$O at redder wavelengths.  Both act to lower the Bond albedo of an
Earth-like planet around a cooler star relative to a hotter star
\citep{Kasting1993}.

\citet{Kopparapu2013} re-calculated the irradiance boundaries using a
\emph{cloud-free} climate model based on new H$_2$O and CO$_2$
absorption coefficients.  The revised boundaries are 10\% lower
(further out) than those of \citet{Selsis2007a} but this difference is
much smaller than that between the cloud-free and cloudy cases of
\citet{Selsis2007a}.  Because the \emph{Kepler} survey is heavily
biased towards shorter periods \citep{Gaidos2013} and thus the
high-irradiance (inner) edge of the HZ is more important to the
determination of $p_{HZ}$, and because of the importance of clouds to
this boundary, I elected to use the 50\% cloud case of
\citet{Selsis2007a}.

I determine whether a planet is in the HZ for each set of model
stellar parameters $\hat{M_*}$,$\hat{L_*}$, and $\hat{T_*}$ with
associated probability $P_i$.  The probability $p_{HZ}$ that the
planet is in the HZ is then:
\begin{equation}
p_{HZ} = \frac{\displaystyle\sum_{i \in HZ}P_i}{\displaystyle\sum_i P_i}
\end{equation}
I consider candidate planets (or planets that may have
satellites) as having greater-than-even odds of orbiting in
the HZ ($p_{HZ} > 0.5$) as well as having a most probable value of
$\tilde{I}$ (with highest $P$) satisfying $I_{out} < \tilde{I} <
I_{in}$.

\section{Results}

\subsection{Comparison with spectroscopic parameters \label{subsec.compare}}

\begin{figure}
\epsscale{0.7}
\plotone{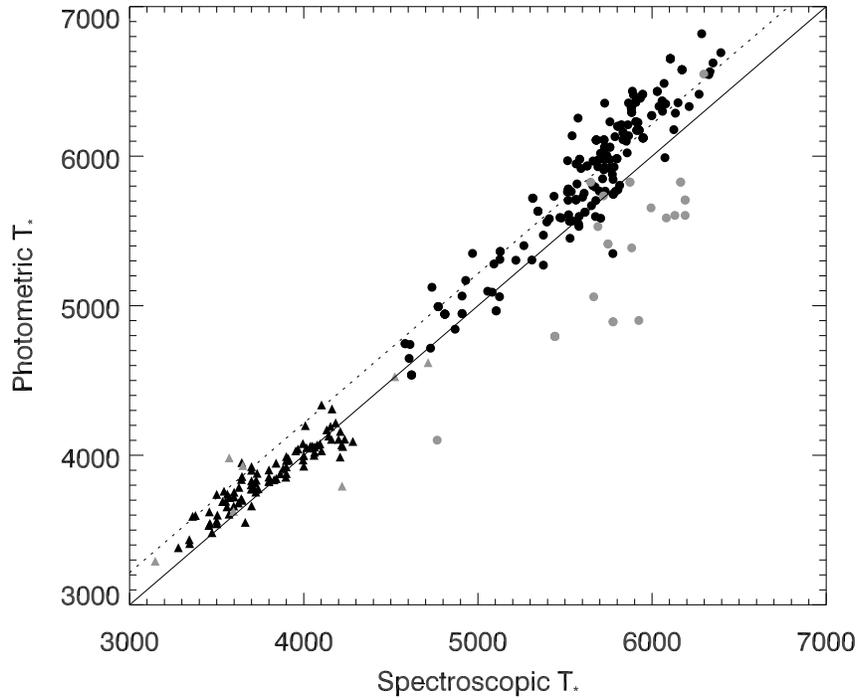}
\caption{Comparison between effective temperatures based on photometry
  and spectroscopic values.  Circles are solar-type stars from
  \citet{Buchhave2012} and \citet{Bruntt2012}.  Triangles are M dwarfs
  from Mann et al., in prep..  Black points are stars where all six
  photometric colors are available; grey points represent stars where
  at least one color is unavailable.  The solid line is equality
  between the estimates and the dashed line represents the $\sim$215~K
  offset found by \citet{Pinsonneault2012}. \label{fig.compareteff}}
\end{figure}

\begin{figure}
\epsscale{0.7}
\plotone{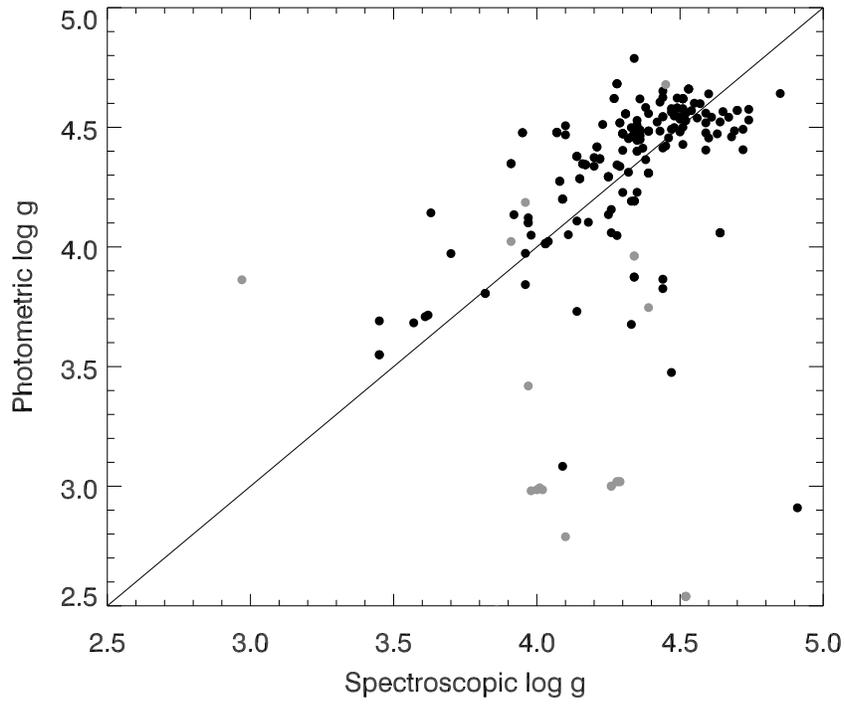}
\caption{Comparison between photometric and spectroscopic estimates of
  surface gravities where the latter are taken from
  \citet{Buchhave2012} and \citet{Bruntt2012}.  Black points are stars
  where all six photometric colors are avialable; grey points
  represent stars where at least one color is unavailable.  The solid
  line is equality between the two estimates.}
  \label{fig.comparelogg}
\end{figure}

Accurate estimates of stellar effective temperature $T_*$ and radius
$R_*$ are crucial to assessing whether a planet is in the HZ, as
together these largely determine the luminosity of the host star and
the irradiance experienced by the planet on a given orbit.  The
inferred radius of a transiting planet also scales linearly with the
estimated radius of the host star.  The radii of distant \emph{Kepler}
stars cannot be directly measured, but spectroscopic values of $T_*$
and surface gravity $\log g$, the latter related to $R_*$, are
available for some \emph{Kepler} stars with planets \citep[][Mann et
al., in prep.]{Bruntt2012,Buchhave2012}.  Figures
\ref{fig.compareteff} and \ref{fig.comparelogg} compare
photometry-based values of $T_*$ and $\log g$ with reported
spectroscopic values.  Photometric values for solar-type stars where
all 6 colors are available average 208~K higher than spectroscopic
values (Fig. \ref{fig.compareteff}).  \citet{Pinsonneault2012} found
that both the original KIC temperatures \emph{and} spectroscopic
estimates were $\sim$215~K cooler than determinations using the
infrared flux method (IRFM).  Thus the new photometric estimates are
in line with IRFM values.  The offset between photometric and
spectroscopic temperatures is less (60~K) for M~dwarfs; spectroscopic
temperatures for these stars (Mann et al., in prep.)  were determined
by comparing spectra to synthetic spectra from PHOENIX/BT-SETTL models
\citep{Allard2011} and tuning the comparison using the temperature
estimates of \citet{Boyajian2012ApJ757}.

Photometric $T_*$ for 16 stars is significantly lower than
spectroscopic estimates.  All but two of these are missing either $i$-
or $z$-band photometry, or both.  The importance of these bandpasses
is not surprising as they are the only source of information in the
wavelength range $0.7 \mu{\rm m} < \lambda < 1.1 \mu{\rm m}$, just
beyond the peak in emission from most of these stars, a spectral
feature which most strongly constrains $T_*$.  There are four stars
with significantly ($>2\sigma$) hotter photometric estimates of $T_*$
relative to spectroscopy; only one of these is missing photometry.
The reason(s) for the discrepancy among the other stars are unclear.
One possibility is that the photometric source is a blend resolved by
spectroscopy, or that the transit signal itself may be coming from a
component of a blend which is dissimilar to the source of most of the
light, and consequently the transit duration prior is skewing the
stellar parameters.  After removing the 208~K offset and ignoring
stars with missing colors, the standard deviation between photometric
and spectroscopic values of $T_*$ is $\sigma = 180$~K for solar-type
stars and 130~K for M dwarfs.  This equals the performance of the
analysis of \citet{Bailer-Jones2011}, but without the benefit of
parallaxes.

Photometry-based estimates of $\log g$ are more discrepant with
spectroscopic values, although an overall correlation is apparent
(Fig. \ref{fig.comparelogg}).  About half of the most discrepant cases
lack photometry in at least one bandpass, although many stars with
missing photometry are assigned surface gravities close to the
spectroscopic estimates.  Although photometric colors involving the
SDSS $u$ \citep{Lenz1998} and $z$ \citep{Vickers2012} bands can be
used to discriminate between hotter main sequence and evolved stars,
photometry is a much blunter tool to separate solar-type stars by
luminosity class.  While my analysis may only marginally improve this
situation, it does quantify the uncertainties.

Among the KOI host stars with reported spectroscopic parameters are
those with candidate HZ planets discussed below (Section
\ref{subsec.habitable}).  \citet{Buchhave2012} report spectroscopic
parameters for three stars in Table \ref{tab.candidates}, including
\emph{Kepler}-22b.  The photometric values of $T_*$ are within 300~K
of the corresponding spectroscopic estimates
(Fig. \ref{fig.compareteff}).  \citet{Muirhead2012} obtained $K$-band
spectra for eight of these HZ stars and Mann et al. (in prep.)
obtained visible-wavelength spectra for 18, including six of the
\citet{Muirhead2012} targets (Table \ref{tab.candidates}).  Spectra
confirm that all 20 are late K- or early M-type dwarfs.  In general,
the photometric temperatures of M dwarf KOI hosts agree with
spectroscopic values except for the case of KIC~10027323 (hosting
KOI~1596.02), where the photometric estimate (4636~K) is 800~K hotter
than an IR spectroscopic value from \citet{Muirhead2012}.  The
\citet{Muirhead2012} temperature are based on H$_2$O indices which
saturate at temperatures hotter than $\sim$3800~K (Mann et al., in
prep.)

\subsection{Planets in the Habitable Zone \label{subsec.habitable}}

Of the 2740 confirmed and candidate planets, the analysis of 1 star
(KIC 7746948 hosting KOIs 326.01 and 326.02) failed, as it is missing
an $r$ magnitude and therefore cannot be analyzed by this procedure.
The majority of (candidate) planets have essentially zero $p_{Hz}$ and
2604 (95\%) have $p_{HZ} < 0.01$.  

\begin{figure}
\epsscale{0.7}
\plotone{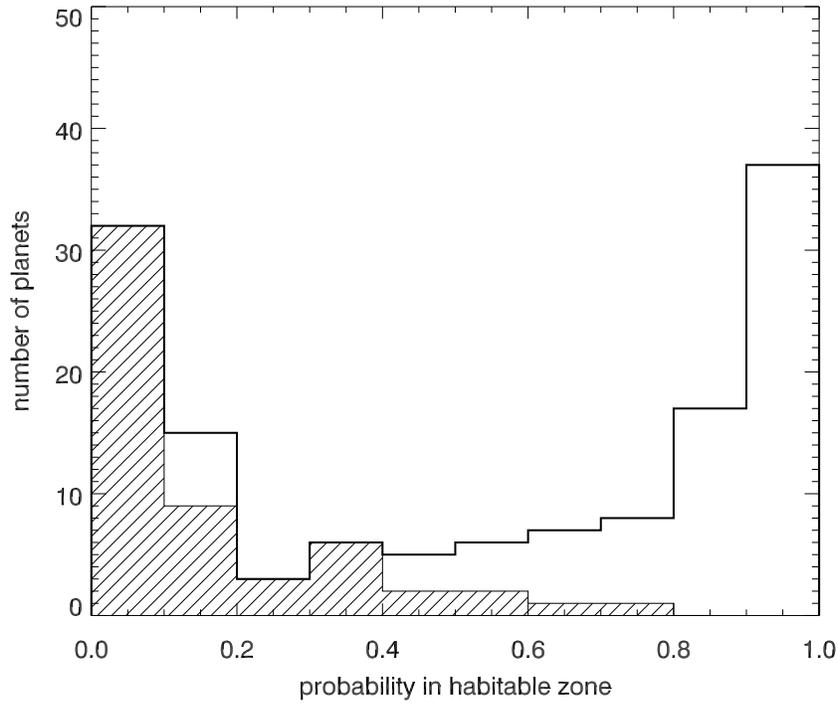}
\caption{Distribution of $p_{HZ}$, the probability that a
  \emph{Kepler} confirmed or candidate planet orbits in its host
  star's habitable zone.  For clarity, only the 136 planets with
  $p_{HZ} > 0.01$ are shown.  The filled histogram is the $p_{HZ}$
  distribution of the subset of objects with maximum posterior
  probability (best-fit) irradiances \emph{outside} the habitable
  zone.}
  \label{fig.hzhist}
\end{figure}

Figure \ref{fig.hzhist} shows the $p_{HZ}$ distribution of the 136
objects with $p_{HZ} > 0.01$; the low-probability tail was excluded
for clarity.  The distribution is qusi-bimodal because some planets
have posterior irradiance PDFs that are narrower than the irradiance
difference across the HZ and hence are either very likely to be ``in''
($p_{HZ} \approx 1$) or ``out'' ($p_{HZ} \approx 0$) of the HZ.  The
expected number of HZ planets in the catalog, the sum of $p_{HZ}$, is
$\sim 73$.  This figure does not change if $p_{HZ} < 0.01$ are
included, i.e. it is not determined by a very large number of low
$p_{HZ}$ objects.  Also plotted in Fig. \ref{fig.hzhist} is the subset
of planets which the maximum posterior probability (best-fit) models
place \emph{outside} the HZ.  These cases arise when stellar
parameters are poorly constrained; all but four have $p_{HZ} < 0.5$,
and I adopted this combination of criteria for identification of the
highest-ranking HZ planets.

Two objects were excluded because they are unlikely to be planets: The
radius of KOI~113.01 is between 1.3\rjupiter{} and 0.37\rsun{} with
95\% confidence and \citet{Batalha2013} list this KOI as having a
``V-shaped'' transit lightcurve indicative of an eclipsing binary.
KOI~1226.01, has a minimum radius of 2\rjupiter{} and a light curve
suggestive of an eclipsing binary \citep{Dawson2012}.  For seven
candidates there is a $>10$\% probabiity that the radius exceeds the
theoretical upper limit for cool Jupiters \citep[$R_p \approx
1.2$\rjupiter{},]{Fortney2010}.  All of these cases could be explained
by the very large errors in the radius of the host star, i.e. the
inability of photometry to rule out an evolved star.  Eight HZ
candidates (KOIs 375.01, 422.01, 435.02, 490.02, 1096.01, 1206.01, and
1421.01) were excluded because their reported orbital periods are
being based on the duration of a single transit and the assumption of
a circular orbit, and have large uncertainties.  

\begin{figure}
\epsscale{0.7} 
\plotone{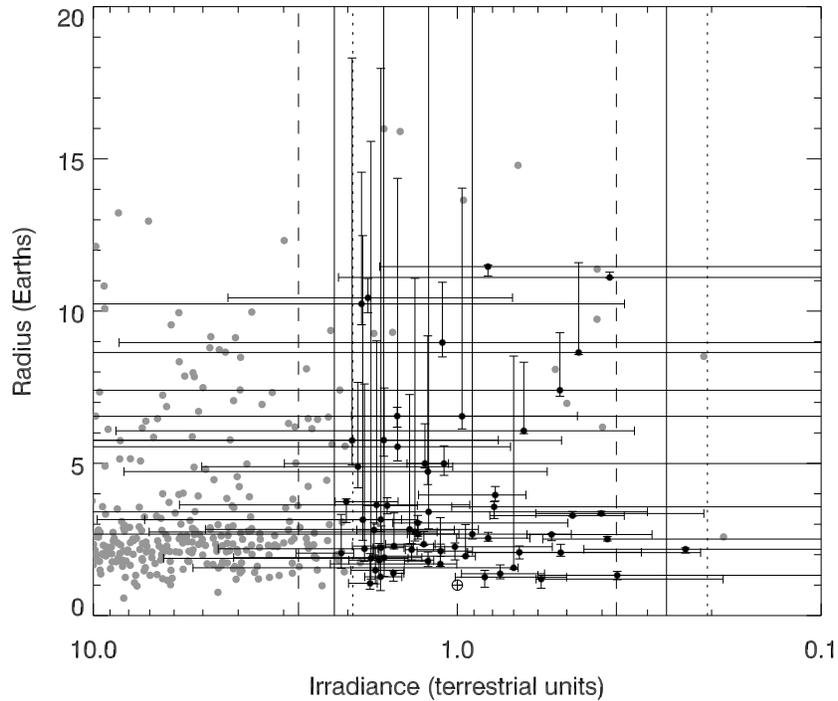}
\caption{Radius and stellar irradiance of candidate and confirmed
  \emph{Kepler} planets, and the Earth.  Candidate planets in
  habitable zones are highlighted as black, all other KOIs are grey,
  and the vast majority of KOIs experience higher irradiances and fall
  outside the left-hand boundary of the plot.  The error bars
  correspond to 95\% confidence intervals.  The solid, dotted, and
  dashed lines are the boundaries of the HZ for a 50\% cloud-covered
  Earth like planet around a solar-type star (5780~K), an early M
  dwarf (3700~K) and late A-type (7800~K) star \citep{Selsis2007a}.}
\label{fig.hz1}
\end{figure}

\begin{figure}
\epsscale{0.7}
\plotone{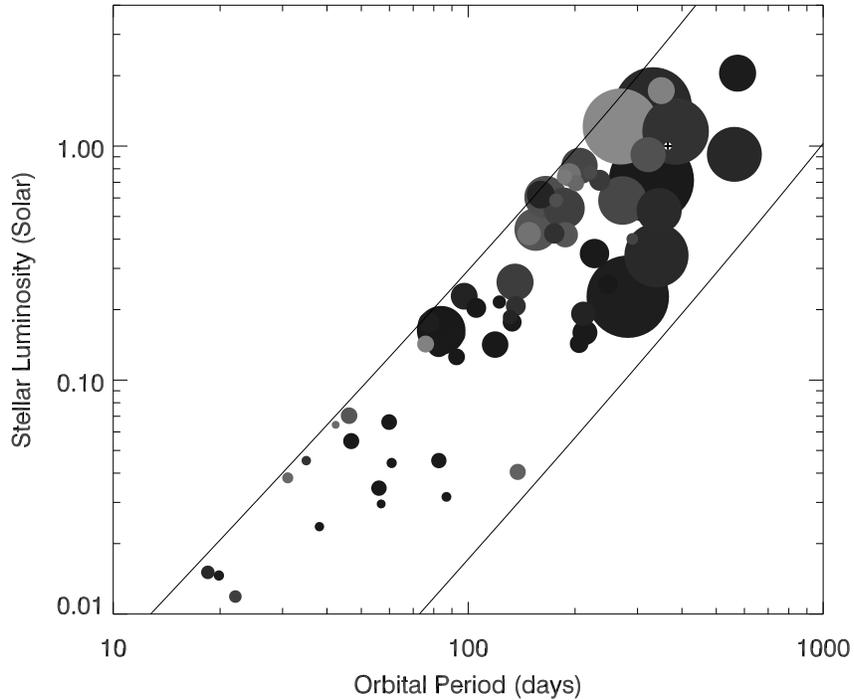}
\caption{Luminosity of the host star vs. orbital period of candidate
  HZ planets ($p_{HZ} > 0.5$) detected by \emph{Kepler}, plus the
  Earth.  The points are scaled to planet radius and the the darker
  the point, the more likely it is in the HZ.  The two lines delimit
  the boundaries of the HZ for Earth-like planets with 50\% cloud
  cover \citep{Selsis2007a}.  To plot the boundaries with these axes,
  it was necessary to assume simple but standard power-law relations
  between the luminosities, masses, and effective temperatures of
  main-sequence stars.}
  \label{fig.hz2}
\end{figure}

\begin{figure}
\epsscale{0.7}
\plotone{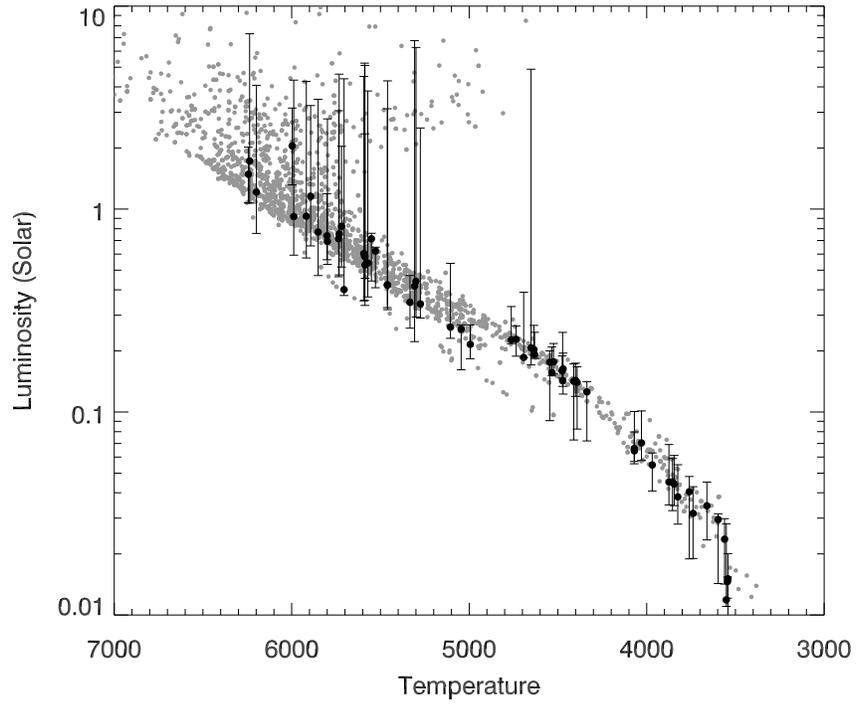}
\caption{Hertzsprung-Russell diagram for host stars of 2739 candidate
  and confirmed \emph{Kepler} planets.  Black points are the 62
  candidate HZ planets in Table 1.  The error bars represent 95\%
  confidence intervals on luminosity and are shown only for the
  candidate HZ planets.}
  \label{fig.hr}
\end{figure}

The 62 remaining candidates with $p_{HZ} > 0.5$ and most probable
incident stellar irradiation in the HZ limits are listed in Table
\ref{tab.candidates} and plotted in Figs. \ref{fig.hz1} and
\ref{fig.hz2}.  The most probable and 95\% confidence intervals for
their irradiance and radius are given, and the stellar parameters of
the model with highest posterior probability are reported.  Figure
\ref{fig.hr} plots the host star parameters in a Hertzsprung-Russell
diagram that includes all 2035 KOI host stars.  Luminosities for a few
host stars have very high upper bounds because the combination of
photometry and priors cannot rule out the possibility that they are
evolved with 95\% confidence.  All are most likely to be dwarfs except
for KOI~1574.02, which I calculate has a probability of 53\% of having
$\log g < 4.2$.  Nearly all are assigned subsolar posterior
metallicities but this is a result of the prior (Section
\ref{sec.priors}) because photometry offers little constraint on
metallicity.

Thirty-four planets were previously identified as possible HZ planets
by \citet{Borucki2011ApJ736}, \citet{Kaltenegger2011}, \citet{Batalha2013},
or \citet{Dressing2013}.  Most, but not all, of the others are
candidates from the January 2013 release.  The candidate around the
brightest host star, KOI~87.01/\emph{Kepler}-22b was previously
flagged by \citet{Borucki2011ApJ736} and \citet{Kaltenegger2011} and
confirmed by \citet{Borucki2012}.  The photometric estimate of
effective temperature (5735~K) is consistent with two spectroscopic
estimates (5518 and 5642~K), the inferred (maximum posterior
probability) luminosity is slightly higher 0.9$L_{\odot}$ compared to
0.79$L_{\odot}$, and the inferred age of 8~Gyr is consistent with slow
rotation and low flux in the core of the Ca II H and K lines
\citep{Borucki2012}.  The inferred stellar mass is identical
(0.93$M_{\odot}$) to that determined by astroseismology.  The
preferred planet radius is 2.32\rearth{} and is within the errors of
the previously published value of $2.38\pm 0.13$\rearth{}, although
the 95\% confidence interval for this star is large.

KOI~250.04 is not (yet) a confirmed planet but is the outermost known
member of the 4-planet \emph{Kepler}-26 system containing two
components (b and c) confirmed by transit timing variation (TTV)
analysis \citep{Steffen2012} and a fourth candidate (KOI 250.03) on
the innermost orbit.  The orbital period of KOI 250.04 ($P_K = 46.83$
d) is suspicously close to one half of the period of a TTV signal seen
near 90~d \citep{Steffen2012}.  This analysis indicates that the host
star of these planets has $T_* = 4072$~K, i.e. is a late K dwarf.
This is confirmed by two moderate-resolution visible-wavelength
spectra which return 3996~K and 4067~K and a spectral type of K7.5
(Mann et al. in prep.), and an infrared spectrum which gives $T_*
=3887$K \citep{Muirhead2012}.  \citet{Steffen2012} report $T_* =
4500$~K based on an SME analysis \citep{Valenti1996} of a Keck-HIRES
spectrum.  However, SME effective temperatures are unreliable for very
cool stars such as this.  KOI 250.04 has a radius of about 2.4\rearth{},
and it is of particular interest because further TTV analysis might
constrain its mass.

The distribution with radius among these candidate HZ planets peaks in
the super-Earth range ($\sim 2.5$\rearth{}) and decreases with
increasing radius, although there may be a cluster of candidates with
radii approximately that of Jupiter.  Presumably gas giants, these
objects are potential hosts for habitable satellites
\citep{Kipping2009,Kaltenegger2010}.  For seven candidates there is a
$>10$\% probabiity that the radius exceeds the theoretical upper limit
for cool Jupiters \citep[$R_p \approx 1.2$\rjupiter{},]{Fortney2010}.
In 5 of these cases, this can be explained by the very large errors in
the radius of the host star (an evolved star cannot be ruled out).  

Most of the smaller planets orbit the lowest-luminosity stars
(Fig. \ref{fig.hz2}), presumably because smaller planets are easier to
detect around smaller stars.  KOIs 2626.01 and 3010.01 are arguably
the most ``Earth-like'' in terms of radius and irradiance.  Table
\ref{tab.candidates} also includes 10 additional candidate planets
with $R_p < 2$\rearth{} and $p_{HZ} > 0.01$ (but $<0.5$).  Five of
these orbit late K or early M-type dwarfs, a figure that supports
claims that these stars are the most promising locales to find
Earth-size and Earth-like planets \citep{Dressing2013}.

\subsection{Not-so-habitable planets}

\citet{Kaltenegger2011} list 27 planets with semimajor axes between
the inner edge (as defined by the onset of a runaway greenhouse) and
outer edge of the HZ.  Of these, 7 (KOIs 113.01, 465.01, 1008.01,
1026.01, 1134.02, 1168.01, 1232.01, were not retained in the
\citet{Batalha2013} catalog.  KOIs~113.01 and 1008.01 have V-shaped
transit shapes and KOI~1232.01 has a large radius indicative of an
elipsing binary.  KOI~1134.02 exhibits ``active pixel offset'' meaning
that the target star is not the source of the transit signal.
KOI~1026.01 might be an artifact of systematics in the \emph{Kepler}
data \citep{Batalha2013}.  KOIs 465.01 and 1168.01 were detected only
with a single transit in the \citet{Borucki2011ApJ736} catalog.  Of the
remaining 20, five (KOIs 139.01, 1099.01, 1423.01, 1439.01, and
1503.01) have $p_{HZ} < 0.5$ and so do not appear in this catalog,
although KOI-1423.01 is omitted marginally only so (0.47).
KOI~1439.01 is most strongly ruled out ($p_{HZ} = 0.06)$ because the
revised $T_*$ is 274~K hotter and $R_*$ is 46\% larger than the KIC
values used by \citet{Kaltenegger2011}.  The other 15 KOIs are
retained in this catalog, along with 5 others from
\citet{Kaltenegger2011}.

A comparison with the HZ candidates of \citet{Batalha2013} is
problematic because they use an equilibrium temperature criterion
which is dependent on the color/effective temperature of the host
star.  However, of the 24 candidate planets with 185K $< T_{eq} <$300K
in Table 8 of \citet{Batalha2013}, one KOI was later eliminated as a
false positive (2841.01), and six KOIs (119.02, 438.02, 986.02,
1938.01, 2020.01, and 2290.01) have $p_{HZ} \ll 0.5$ and/or most probable
$<I>$ outside the HZ limits.  In each case, this is because the new
estimates for $T_*$ are $\ge$200~K hotter than the previously
published values, and because the most probably estimate of radius is
significantly larger.

\subsection{Fraction of \emph{Kepler} stars with planets in the Habitable Zone}

The calculations described above can be applied to the entire
\emph{Kepler} target catalog to estimate the fraction $f_{HZ}$ of stars
with planets orbiting in the habitable zone.  Obviously, the
constraint on stellar density from the durations of transits could
only be applied to KOIs.  I estimated $f_{HZ}$ using the detection
statistics of planets with $P < 245$~d (at least 3 transits over 2~yr)
around 122,442 stars with $\log g > 4$ (KIC value\footnote{KIC $\log
  g$ is sometimes unreliable, but is usually an overestimate, and thus
  few dwarf stars are excluded.}) observed for at least 7 quarters of
Q1-8.

This calculation identified the value of $f_{HZ}$ that maximizes the
logarithmic likelihood \citep[e.g.,][]{Mann2012}
\begin{equation}
\label{eqn.hzprob}
\ln L = \displaystyle \sum_i^D \ln \left(f_{HZ}\langle d_{ik}\rangle \right) + \sum_j^{ND} \ln \left(1-f_{HZ}\langle d_{jk}\rangle \right),
\end{equation}
where the first and second sums are over systems with and without
detected planets with $R_p > 2$\rearth{} and $P < 245$~d in the HZ,
respectively, $d_{ik}$ is the probability of detecting a planet in the
HZ of the $i$th star described by the $k$th model, and $\langle
\rangle$ represents the weighted average over all relevant models.

The detection completeness of the \emph{Kepler} survey for planets wth
$R_p < 2$\rearth{} is still being established.  I estimated $\eta$ for
$R_p > 0.8$\rearth{} by first computing the value for $R_p >
2$\rearth{}, then adjusting by the ratio 2.5 of $R_P > 0.8$\rearth{}
to $R_p > 2$\rearth{} planets with $P < 85$~d planets based on Table 3
of \citet{Fressin2013}.  This manuever assumes that the planet
population inside 85~d is the same as that inside 245~d, but a
distribution with radius would have to be assumed regardless because
of severe incompleteness for small planets on wider orbits.

I calculated $d$ as the product of the geometric probability of
transiting $d_{\rm transit}$, averaged over the HZ, and the fraction
of planets $d_{signal}$ with $R_P > 2$\rearth{} that would produce a
transit large enough to be detected.  For planets on circular orbits
that are log-distributed with $P$ by a power-law with index $\beta$,
the orbit-averaged geometric detection probability is:
\begin{equation}
d_{\rm transit}  = 0.00465 \left(\frac{\hat{\rho}_*}{\rho_{\odot}}\right)^{-1/3}\left(\frac{P_{\rm in}}{\rm 1~yr}\right)^{-2/3}\frac{\beta}{\beta + \frac{2}{3}}\frac{1-\left(P_{\rm in}/P_{\rm out}\right)^{\beta + 2/3}}{1 - \left(P_{\rm in}/P_{\rm max}\right)^{\beta}},
\end{equation}
where $P_{\rm in}$, $P_{\rm out}$, and $P_{\rm max}$ are the orbital
periods at the inner and outer edges of the habitable zone and either
the outer edge or the maximum period of the survey (245~d),
respectively.  These also depend on the luminosity and mass of the
stellar model.  

\begin{figure}
\epsscale{0.7}
\plotone{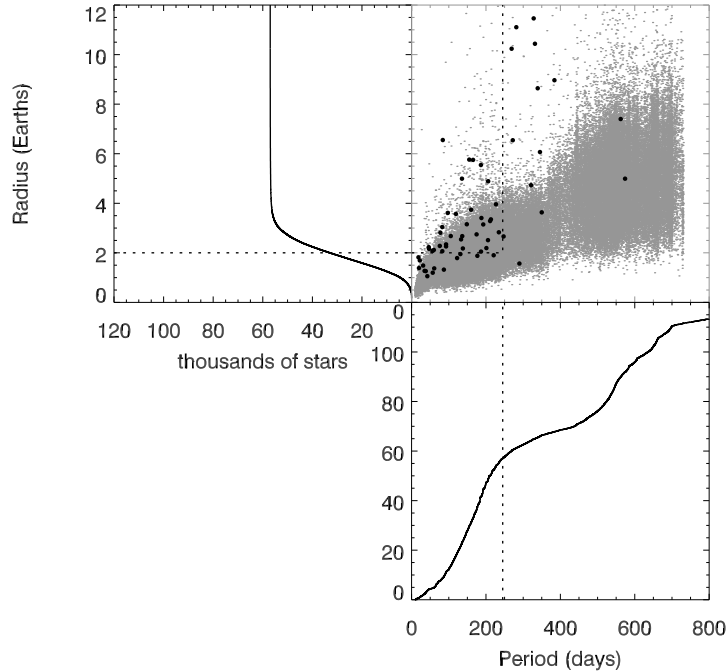}
\caption{Upper right: scatter plot of minimum detectable planet radius
  $R_{\rm min}$ vs. orbital period at the inner edge of the habitable
  zone $P_{\rm in}$ for 122,442 \emph{Kepler} stars observed for at
  least seven of Quarters 1-8 (grey points).  Black points are the
  candidate HZ planets listed in Table \ref{tab.candidates}.  Bottom
  right: cumulative distribution with $P_{\rm in}$.  Upper left:
  cumulative distribution with $R_{\rm min}$ for stars with $P_{\rm
    in} < 245$~d.  Dashed lines indicate the boundaries used to
  calculate the fraction of stars with planets in the habitable zone.}
  \label{fig.hzfrac}
\end{figure}

Based on a power-law distribution wth log radius \citep{Howard2012},
the fraction of planets $d_{\rm signal}$ generating a detectable
transit was taken to be $(R_{\rm min}/2$\rearth$)^{-1.92}$, if $R_{\rm
  min} > 2$\rearth{}, or unity otherwise.  $R_{\rm min}$ is the
minimum radius for detection (SNR = 7.1):
\begin{equation}
  R_{\rm min} = 0.29R_{\oplus}\frac{\hat{R}_*}{R_{\odot}} \left[{\rm CDPP}_6\sqrt{\frac{\rm 6~hr}{DN}}\right]^{1/2},
\end{equation}
where CDPP$_6$ is the average 6~hr Combined Differential Photometric
Precision over Q1-8 (in ppm), $D$ is the transit duration at the inner
edge of the HZ, and $N$ is the number of transits in 2~yr for a planet
with $P_{\rm in}$.  Figure \ref{fig.hzfrac} shows a scatterplot and
cumulative distributions of $P_{\rm in}$ and $R_{\rm min}$ for all
stars assessed for these calculations.  Forty-eight planets and
$\sim$57,000 stars actually contributed to the statistics.

The presence of a planet in the HZ is known only with confidence
$p_{HZ}$.  To account for this, 10,000 Monte Carlo realizations of
detections and non-detections were generated using the values of
$p_{HZ}$ for each star, specifically new values of $p_{HZ}$ which
represent the probability of a planet in the HZ having $R_p >
2$\rearth{}.  The probability distributions with $f_{HZ}$
(Eqn. \ref{eqn.hzprob}) were computed for each realization and summed.
The summed distribution peaks at 0.332 with 95\% confidence limits of
0.22 and 0.49.  Based on the distribution in \citet{Fressin2013}, the
fraction of stars with a planet larger than 0.8\rearth{} in the HZ is
$1-(1-0.332)^{2.5} = 0.64$ (95\% confidence interval of 0.46-0.81).

\section{Discussion \label{sec.discussion}}

{\it Assumptions and systematic errors:} There are several
approximations and potential sources of systematic error that could
affect the values of $p_{HZ}$ calculated here; I expect these values
to evolve and that a few candidate planets may move in or out of the
catalog as new data are incorporated, and the DSEP models are revised.
However, the close correspondence between this catalog and previous
ones suggests that the selection is relatively robust, although the
relative rankings may change.

The constraint from the transit duration depends on orbital
eccentricity, argument of periastron, and impact parameter.  Uniform
priors are appropriate choices for the last two parameters.  However,
a Rayleigh distribution for eccentricities
(Eqn. \ref{eqn.eccentricity}) with mean $\langle e \rangle = 0.2$,
while consistent with \emph{Kepler} data \citep{Moorhead2011a}, is
neither tightly constrained nor a unique choice
\citep[e.g.][]{Shen2008}.  Indeed, a more refined prior would include
the interrelationships with planet mass, orbital period, and the age
of the system \citep{Wang2011}.  I calculated the difference in
$p_{HZ}$ resulting from changing $\langle e \rangle$ from 0.1 to 0.3.
For the 62 candidates in the HZ, one half of the mean difference
between the $p_{HZ}$ values is 0.019.  This indicates that the transit
duration constraint has a small but non-negligible effect on the
identification of HZ planets.

These priors do not include the probability that a planet will transit
its host star and be detected by \emph{Kepler}, and thus be included
in the KOI catalog.  Such selection effects can be important in
catalogs of transiting planets and their host stars
\citep{Gaidos2013}.  The geometric transit probability $R_*/a$, where
$a$ is the semimajor axis, is proportional to $\tau^{2/3}$ and could
be included readily enough: this factor will favor stellar models with
larger radii.  However, the probability of transit detection is
primarily related to transit depth $\delta \approx (R_p/R_*)^2$ and
for a given $\delta$, a prior on stellar radius is ultimately a prior
on \emph{planet} radius.  Some of these KOIs are nearly Earth-size,
where the completeness of the \emph{Kepler} survey is still being
refined.  Other KOIs are at or near theoretical limits of giant planet
radii and any prior on stellar radii would have to include scenarios
for astrophysical false positives.  There are additional, but perhaps
minor complexities: the probability of a transit occurring and being
detected will also depend on $e$, $\phi$, $b$, as well as $D$, the
transit duration.  For these reasons, I do not include transit
detection as a prior.

Equation \ref{eqn.reddening} presumes a linear relationship between
extinction in different bandpasses, i.e. that all can be linearly
related to reddening $E_{B-V}$.  This is not strictly correct, but is
a fair approximation in the limit of small reddening.  The median
derived $E_{B-V}$ for these stars is only 0.08, corresponding to 0.25
magnitudes of extinction, and the 95 percentile value is 0.18.  If the
scale height of dust is smaller than that of stars, then the uniform
prior derived under the assumption of identical gas and dust
distributions (Appendix A) slightly underestimates the amount of
reddening.  Because reddening and temperatures derived from photometry
are correlated, this assumption slightly underestimates the
temperature and luminosities of stars as well.

The total number of candidate HZ planets is not sensitive to the
precise irradiation limits.  Because of detection bias towards
short-period orbits, there are very few detected planets beyond the HZ
(Fig. \ref{fig.hz1}).  For an Earth-like planet with 100\% cloud
cover, the runaway greenhouse irradiation limit is 23\% higher than
the 50\% cloud-cover case \citep{Selsis2007a}, but this admits only
one additional candidate to the catalog.  On the other hand, HZ
calculations are sensitive to the precise value of $T_*$ because of
the sensitivity of luminosity to effective temperature, and future
refinements are worthwhile (see below).  I do not account for
systematic errors in the DSEP and TRILEGAL models themselves, but
given the agreement with spectroscopy (Fig. \ref{fig.compareteff})
these are likely to be comparatively small.  Of course, the HZ
described here only applies to Earth-like planets with a surface
pressure of $\sim1$~bar.  Planets with different surface gravities,
pressures, and/or compositions may be habitable to larger distances
\citep{Pierrehumbert2011ApJ734}, or not at all \citep{Gaidos2000}.

{\it The trouble with M dwarfs:} The fundamental parameters of M dwarf
stars have been a notorious challenge for models because of the
difficulty in reproducing the observed mass-radius relation and their
complex spectra.  The DSEP models employed here accurately predict the
radii of the two M dwarfs in the triply eclipsing hierarchical triple
system KOI-126 \citep{Feiden2011} \citep[see also][]{Feiden2012}.
DSEP uses PHOENIX model atmospheres \citep{Hauschildt1999} for both
the stellar surface boundary conditions and to generate synthetic
magnitudes.  The spectroscopic temperatures presented here are
calibrated using nearby interferometry targets
\citep{Boyajian2012ApJ757} using the BT-SETTL flavor of PHOENIX models
\citep{Lepine2013}, hence the good correlation between the two
estimates is not surprising.  Nevertheless, the offset of 60~K in
$T_*$ is represents a $\sim$10\% difference in $L_*$.

Another obstacle is that accurate modeling of the lightcurves of
planets transiting M dwarfs must correctly account for significant
limb darkening in the \emph{Kepler} pass-band.  Erroneous transit
durations, acting through the prior described in Section
\ref{sec.transit}, can bias the analysis towards models with incorrect
radii: a 10\% error in $R_*$ leads to a $\sim$30\% error in $L_*$.
This is sufficient to ``move'' a planet completely outside the HZ, or
at least decrease the $p_{HZ}$ of a marginal HZ planet to $<$50\%.
Re-analyses of the \emph{Kepler} transit lightcurves with improved
limb-darkening models and re-derivation of the parameters of M dwarf
KOI hosts are worthwhile, \citep[e.g.,][]{Dressing2013}.

{\it Future observations of Kepler stars:} Stellar parameters based on
analysis of photometry are no substitute for values based on
high-resolution spectra, as long as the latter are carefully
calibrated \citep[see][]{Pinsonneault2012}.  However, the median
magnitude of the host stars of these planet is $K_p \approx 15.1$, and
high-resolution spectroscopy is observationally expensive.  The object
most amenable to followup is, not coincidentally, \emph{Kepler}-22b
($K_p = 11.7$).  The next brightest host star is that of KOI~1989.01
($K_p = 13.3$) and the rest are much fainter still and would require
significant time on very large telescopes.  However, this analysis
generates a robustly-defined catalog to prioritize such work.

The \emph{Gaia} (originally Global Astrometric Interferometer for
Astrophysics) mission, scheduled for launch in October 2013, will
obtain parallaxes with a sky-averaged, end-of-mission precision of
25~$\mu$as and 40~$\mu$as for 15th and 16th magnitude stars,
respectively, and somewhat superior performance at the ecliptic
latitude ($\sim$66~deg.) of the \emph{Kepler} field
\citep{deBruijne2012}.  To assess the potential of \emph{Gaia} to
refine the habitable zones of \emph{Kepler} stars and the sizes of the
planest that inhabit them, I re-calculated $P_i$ (Eqn. \ref{eqn.prob})
for all models using a prior for distance modulus $p_{\mu}$ based on
\emph{Gaia}'s expected precision:
\begin{equation}
p_{\mu}= {\rm exp}\:\left[-\frac{\left(\mu - \mu_0\right)^2}{2\sigma_{\mu}^2}\right],
\end{equation}
where $\mu_0$ is the most probable distance modulus from the original
analysis, and $\sigma_{\mu} \approx 8.7\times 10^{\mu_0/5-4}$ is the
uncertainty in $\mu$ from a 40 $\mu$as precision in parallax.

\begin{figure}
\epsscale{0.7}
\plotone{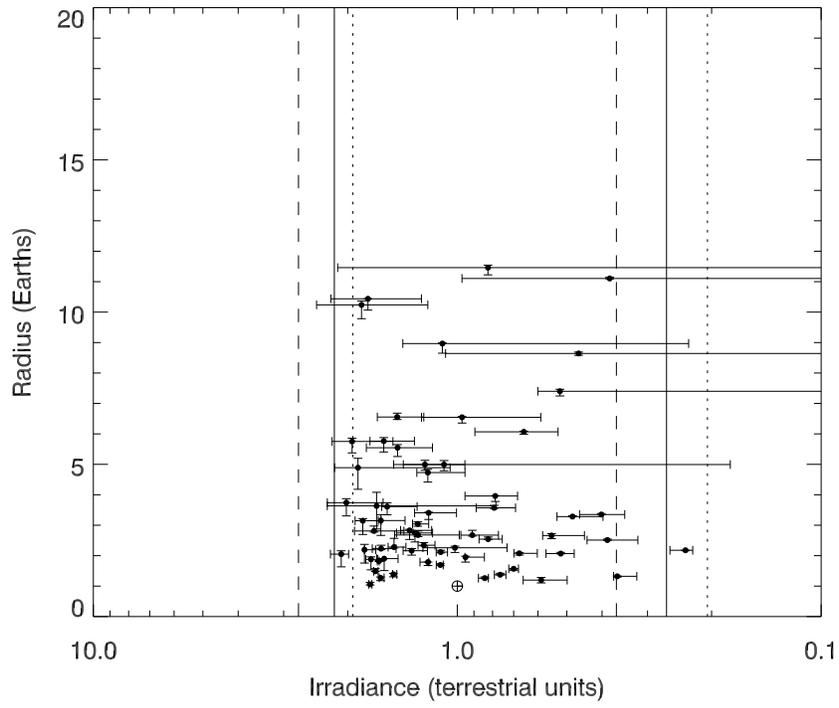}
\caption{Reduction in the uncertainties in planet radius and stellar
  irradiation expected from inclusion of \emph{Gaia} parallax
  measurements with 40~$\mu$as errors.  Compare to
  Fig. \ref{fig.hz1}.}
  \label{fig.gaia}
\end{figure}

The 95\% confidence intervals in radius and stellar irradiance of the
62 HZ candidates were re-calculated and are shown in
Fig. \ref{fig.gaia}.  The most probable values are unchanged, but the
fractional errors in radius and irradiance are reduced by a factor of
$\sim$5, from a median of 10\% and 24\%, respectively, to 1.7\% and
5\%, (equating 95\% confidence intervals to $4\sigma$).  The largest
planets tend to orbit the hottest and most distant stars
\citep{Gaidos2013} and their parameters would retain the largest
errors in this scenario.  Typically, a few hundred DSEP models have
appreciable $P$ values and contribute to the calculation for each
star, but in a few cases the number is a few dozen and finite model
grid size may determine the size of the errors.  Values of $p_{HZ}$
for 50 of the 62 planets are $>0.97$.  Spectroscopic values of $T_*$
accurate to 100~K would offer only modest further improvement (1.5\%
and 4.5\% errors, respectively).  Because these precisions reach or
exceed levels of confidence in the predictions of the stellar models
themselves as well as the absolute calibration of the photometry,
refinement and verification of these may prove a more cost-efficient
avenue for improvement.  For example, \emph{Ugri} and H$_{\alpha}$
photometry of much of the \emph{Kepler} field has been obtained at the
Isaac Newton Telescope \citep{Greiss2012} and \emph{UBV} photometry
has been obtained at WIYN \citep{Everett2012}.

With such precision, it should be possible to locate planets within
different regions of the HZ, e.g. near the inner edge, where low
CO$_2$ atmospheres, and possibly high cloud fraction if there is a
temperature-cloud feedback, should prevail: or the outer edge, where
high CO$_2$ \citep{vonParis2013} and possible water cloud-free
atmospheres are more likely.  Candidate HZ planets in multi-planet
systems might be confirmed or even have masses determined by TTV.
Although such advances may be difficult for planets around faint {\it
  Kepler} stars, this analysis offers a preview of the potential
return from surveys of nearby, more observationally accessible stars,
e.g. by the proposed TESS \citep{Deming2009} and CHEOPS missions.

My calculations suggest that $\sim$64\% of dwarf stars have planets
orbiting in their habitable zones. The fraction of stars with
Earth-size ($R_p = 0.8-2$\rearth{}) planets in the HZ (\etaearth{}) is
0.46 (95\% confidence limits of 0.31-0.64). This statistic will be
greatly refined as the \emph{Kepler} extended mission more thoroughly
probes the HZ of solar-type stars, detection completeness is better
quantified for smaller planets (Fig. \ref{fig.hzfrac}), and the
luminosities of the stars are better established (Fig \ref{fig.gaia}).
This estimate is only marginally higher than that of \citet{Traub2012}
(\etaearth{} = $0.34\pm0.14$), who used the first 136 days of
\emph{Kepler} data.  Also using \emph{Kepler} data,
\citet{Dressing2013} calculated that $0.15^{+0.13}_{-0.6}$ of M dwarfs
have Earth-size (0.5-1.4\rearth{}) planets in the HZ, but this was
revised upwards to $0.48^{+0.12}_{-0.24}$ by \citet{Kopparapu2013b}.
Based on a radial velocity survey, \citet{Bonfils2011} estimated that
0.41$^{+0.54}_{-0.13}$ of M dwarfs have planets with 1\mearth{}$<
M_p\sin i <$10\mearth{} in the HZ.  The latter estimates are
completely consistent with the value reported here for a wider range
of spectral types, supporting optimism that numerous planets orbit in
the habitable zones of stars all along the main sequence.  Setting
aside questions of formation and long-term orbital stability, these
statistics also suggest favorable odds for finding a planet in the HZ
of a component of the nearest star system, $\alpha$ Centauri.

\clearpage

\acknowledgments

This research was supported by NSF grants AST-09-08406 and NASA grants
NNX10AI90G and NNX11AC33G.  The \emph{Kepler} mission is funded by the
NASA Science Mission Directorate, and data were obtained from the
Mukulski Archive at the Space Telescope Science Institute, funded by
NASA grant NNX09AF08G, and the NASA Exoplanet Archive at IPAC.  Andrew
Mann kindly provided stellar parameters in advance of publication.

%\bibliographystyle{/Users/PEBL/Desktop/Bibliography/apj}
%\bibliography{/Users/PEBL/Desktop/Bibliography/apj-jour,/Users/PEBL/Desktop/Bibliography/additional,/Users/PEBL/Desktop/Bibliography/allrefs}

\begin{thebibliography}{68}
\expandafter\ifx\csname natexlab\endcsname\relax\def\natexlab#1{#1}\fi

\bibitem[{Abe {et~al.}(2011)Abe, Abe-Ouchi, Sleep, \& Zahnle}]{Abe2011}
Abe, Y., Abe-Ouchi, A., Sleep, N.~H., \& Zahnle, K.~J. 2011, Astrobiology, 11,
  443

\bibitem[{{Allard} {et~al.}(2011){Allard}, {Homeier}, \&
    {Freytag}}]{Allard2011} {Allard}, F., {Homeier}, D., \& {Freytag},
  B. 2011, in Astronomical Society of the Pacific Conference Series,
  Vol. 448, 16th Cambridge Workshop on Cool Stars, Stellar Systems,
  and the Sun, ed. C.~{Johns-Krull}, M.~K. {Browning}, \&
  A.~A. {West}, 91

\bibitem[{{Ammons} {et~al.}(2006){Ammons}, {Robinson}, {Strader}, {Laughlin},
  {Fischer}, \& {Wolf}}]{Ammons2006}
{Ammons}, S.~M., {Robinson}, S.~E., {Strader}, J., {Laughlin}, G., {Fischer},
  D., \& {Wolf}, A. 2006, \apj, 638, 1004

\bibitem[{{Bailer-Jones}(2011)}]{Bailer-Jones2011}
{Bailer-Jones}, C.~A.~L. 2011, \mnras, 411, 435

\bibitem[{{Batalha} {et~al.}(2010){Batalha}, {Borucki}, {Koch}, {Bryson},
  {Haas}, {Brown}, {Caldwell}, {Hall}, {Gilliland}, {Latham}, {Meibom}, \&
  {Monet}}]{Batalha2010}
{Batalha}, N.~M., {et~al.} 2010, \apjl, 713, L109

\bibitem[{{Batalha} {et~al.}(2013){Batalha}, {Rowe}, {Bryson}, {Barclay},
  {Burke}, {Caldwell}, {Christiansen}, {Mullally}, {Thompson}, {Brown},
  {Dupree}, {Fabrycky}, {Ford}, {Fortney}, {Gilliland}, {Isaacson}, {Latham},
  {Marcy}, {Quinn}, {Ragozzine}, {Shporer}, {Borucki}, {Ciardi}, {Gautier},
  {Haas}, {Jenkins}, {Koch}, {Lissauer}, {Rapin}, {Basri}, {Boss}, {Buchhave},
  {Carter}, {Charbonneau}, {Christensen-Dalsgaard}, {Clarke}, {Cochran},
  {Demory}, {Desert}, {Devore}, {Doyle}, {Esquerdo}, {Everett}, {Fressin},
  {Geary}, {Girouard}, {Gould}, {Hall}, {Holman}, {Howard}, {Howell},
  {Ibrahim}, {Kinemuchi}, {Kjeldsen}, {Klaus}, {Li}, {Lucas}, {Meibom},
  {Morris}, {Pr{\v s}a}, {Quintana}, {Sanderfer}, {Sasselov}, {Seader},
  {Smith}, {Steffen}, {Still}, {Stumpe}, {Tarter}, {Tenenbaum}, {Torres},
  {Twicken}, {Uddin}, {Van Cleve}, {Walkowicz}, \& {Welsh}}]{Batalha2013}
---. 2013, \apjs, 204, 24

\bibitem[{{Belikov} \& {R{\"o}ser}(2008)}]{Belikov2008}
{Belikov}, A.~N., \& {R{\"o}ser}, S. 2008, \aap, 489, 1107

\bibitem[{{Bonfils} {et~al.}(2011){Bonfils}, {Delfosse}, {Udry}, {Forveille},
  {Mayor}, {Perrier}, {Bouchy}, {Gillon}, {Lovis}, {Pepe}, {Queloz}, {Santos},
  {S{\'e}gransan}, \& {Bertaux}}]{Bonfils2011}
{Bonfils}, X., {et~al.} 2011, ArXiv e-prints 1111.5019

\bibitem[{{Borucki} {et~al.}(2010){Borucki}, {Koch}, {Basri}, {Batalha},
  {Brown}, {Caldwell}, {Caldwell}, {Christensen-Dalsgaard}, {Cochran},
  {DeVore}, {Dunham}, {Dupree}, {Gautier}, {Geary}, {Gilliland}, {Gould},
  {Howell}, {Jenkins}, {Kondo}, {Latham}, {Marcy}, {Meibom}, {Kjeldsen},
  {Lissauer}, {Monet}, {Morrison}, {Sasselov}, {Tarter}, {Boss}, {Brownlee},
  {Owen}, {Buzasi}, {Charbonneau}, {Doyle}, {Fortney}, {Ford}, {Holman},
  {Seager}, {Steffen}, {Welsh}, {Rowe}, {Anderson}, {Buchhave}, {Ciardi},
  {Walkowicz}, {Sherry}, {Horch}, {Isaacson}, {Everett}, {Fischer}, {Torres},
  {Johnson}, {Endl}, {MacQueen}, {Bryson}, {Dotson}, {Haas}, {Kolodziejczak},
  {Van Cleve}, {Chandrasekaran}, {Twicken}, {Quintana}, {Clarke}, {Allen},
  {Li}, {Wu}, {Tenenbaum}, {Verner}, {Bruhweiler}, {Barnes}, \&
  {Prsa}}]{Borucki2010Sci}
{Borucki}, W.~J., {et~al.} 2010, Science, 327, 977

\bibitem[{{Borucki} {et~al.}(2011){Borucki}, {Koch}, {Basri}, {Batalha},
  {Brown}, {Bryson}, {Caldwell}, {Christensen-Dalsgaard}, {Cochran}, {DeVore},
  {Dunham}, {Gautier}, {Geary}, {Gilliland}, {Gould}, {Howell}, {Jenkins},
  {Latham}, {Lissauer}, {Marcy}, {Rowe}, {Sasselov}, {Boss}, {Charbonneau},
  {Ciardi}, {Doyle}, {Dupree}, {Ford}, {Fortney}, {Holman}, {Seager},
  {Steffen}, {Tarter}, {Welsh}, {Allen}, {Buchhave}, {Christiansen}, {Clarke},
  {Das}, {D{\'e}sert}, {Endl}, {Fabrycky}, {Fressin}, {Haas}, {Horch},
  {Howard}, {Isaacson}, {Kjeldsen}, {Kolodziejczak}, {Kulesa}, {Li}, {Lucas},
  {Machalek}, {McCarthy}, {MacQueen}, {Meibom}, {Miquel}, {Prsa}, {Quinn},
  {Quintana}, {Ragozzine}, {Sherry}, {Shporer}, {Tenenbaum}, {Torres},
  {Twicken}, {Van Cleve}, {Walkowicz}, {Witteborn}, \&
  {Still}}]{Borucki2011ApJ736}
---. 2011, \apj, 736, 19

\bibitem[{{Borucki} {et~al.}(2012){Borucki}, {Koch}, {Batalha}, {Bryson},
  {Rowe}, {Fressin}, {Torres}, {Caldwell}, {Christensen-Dalsgaard}, {Cochran},
  {DeVore}, {Gautier}, {Geary}, {Gilliland}, {Gould}, {Howell}, {Jenkins},
  {Latham}, {Lissauer}, {Marcy}, {Sasselov}, {Boss}, {Charbonneau}, {Ciardi},
  {Kaltenegger}, {Doyle}, {Dupree}, {Ford}, {Fortney}, {Holman}, {Steffen},
  {Mullally}, {Still}, {Tarter}, {Ballard}, {Buchhave}, {Carter},
  {Christiansen}, {Demory}, {D{\'e}sert}, {Dressing}, {Endl}, {Fabrycky},
  {Fischer}, {Haas}, {Henze}, {Horch}, {Howard}, {Isaacson}, {Kjeldsen},
  {Johnson}, {Klaus}, {Kolodziejczak}, {Barclay}, {Li}, {Meibom}, {Prsa},
  {Quinn}, {Quintana}, {Robertson}, {Sherry}, {Shporer}, {Tenenbaum},
  {Thompson}, {Twicken}, {Van Cleve}, {Welsh}, {Basu}, {Chaplin}, {Miglio},
  {Kawaler}, {Arentoft}, {Stello}, {Metcalfe}, {Verner}, {Karoff}, {Lundkvist},
  {Lund}, {Handberg}, {Elsworth}, {Hekker}, {Huber}, {Bedding}, \&
  {Rapin}}]{Borucki2012}
---. 2012, \apj, 745, 120

\bibitem[{{Boyajian} {et~al.}(2012){Boyajian}, {von Braun}, {van Belle},
  {McAlister}, {ten Brummelaar}, {Kane}, {Muirhead}, {Jones}, {White},
  {Schaefer}, {Ciardi}, {Henry}, {L{\'o}pez-Morales}, {Ridgway}, {Gies}, {Jao},
  {Rojas-Ayala}, {Parks}, {Sturmann}, {Sturmann}, {Turner}, {Farrington},
  {Goldfinger}, \& {Berger}}]{Boyajian2012ApJ757}
{Boyajian}, T.~S., {et~al.} 2012, \apj, 757, 112

\bibitem[{{Brown} {et~al.}(2011){Brown}, {Latham}, {Everett}, \&
  {Esquerdo}}]{Brown2011AJ142}
{Brown}, T.~M., {Latham}, D.~W., {Everett}, M.~E., \& {Esquerdo}, G.~A. 2011,
  \aj, 142, 112

\bibitem[{{Bruntt} {et~al.}(2012){Bruntt}, {Basu}, {Smalley}, {Chaplin},
  {Verner}, {Bedding}, {Catala}, {Gazzano}, {Molenda-{\.Z}akowicz}, {Thygesen},
  {Uytterhoeven}, {Hekker}, {Huber}, {Karoff}, {Mathur}, {Mosser},
  {Appourchaux}, {Campante}, {Elsworth}, {Garc{\'{\i}}a}, {Handberg},
  {Metcalfe}, {Quirion}, {R{\'e}gulo}, {Roxburgh}, {Stello},
  {Christensen-Dalsgaard}, {Kawaler}, {Kjeldsen}, {Morris}, {Quintana}, \&
  {Sanderfer}}]{Bruntt2012}
{Bruntt}, H., {et~al.} 2012, \mnras, 423, 122

\bibitem[{Buchhave {et~al.}(2012)Buchhave, Latham, Johansen, Bizzarro, Torres,
  Rowe, Batalha, Borucki, Brugamyer, Caldwell, Bryson, Ciardi, Cochran, Endl,
  Esquerdo, Ford, Geary, Gilliland, Hansen, Isaacson, Laird, Lucas, Marcy,
  Morse, Robertson, Shporer, Stefanik, Still, \& Quinn}]{Buchhave2012}
Buchhave, L.~a., {et~al.} 2012, Nature, 486, 375

\bibitem[{{Chen} {et~al.}(2007){Chen}, {Yang}, \& {Zhang}}]{Chen2007}
{Chen}, P.-S., {Yang}, X.-H., \& {Zhang}, P. 2007, \aj, 134, 214

\bibitem[{{Cutri} {et~al.}(2003){Cutri}, {Skrutskie}, {van Dyk}, {Beichman},
  {Carpenter}, {Chester}, {Cambresy}, {Evans}, {Fowler}, {Gizis}, {Howard},
  {Huchra}, {Jarrett}, {Kopan}, {Kirkpatrick}, {Light}, {Marsh}, {McCallon},
  {Schneider}, {Stiening}, {Sykes}, {Weinberg}, {Wheaton}, {Wheelock}, \&
  {Zacarias}}]{Cutri2003}
{Cutri}, R.~M., {et~al.} 2003, {2MASS All Sky Catalog of point sources.}

\bibitem[{{Dawson} {et~al.}(2012){Dawson}, {Murray-Clay}, \&
  {Johnson}}]{Dawson2012}
{Dawson}, R.~I., {Murray-Clay}, R.~A., \& {Johnson}, J.~A. 2012, ArXiv e-prints 1211.0554

\bibitem[{{de Bruijne}(2012)}]{deBruijne2012}
{de Bruijne}, J.~H.~J. 2012, \apss, 341, 31

\bibitem[{{Demarque} {et~al.}(2004){Demarque}, {Woo}, {Kim}, \&
  {Yi}}]{Demarque2004}
{Demarque}, P., {Woo}, J.-H., {Kim}, Y.-C., \& {Yi}, S.~K. 2004, \apjs, 155,
  667

\bibitem[{{Deming} {et~al.}(2009){Deming}, {Seager}, {Winn}, {Miller-Ricci},
  {Clampin}, {Lindler}, {Greene}, {Charbonneau}, {Laughlin}, {Ricker},
  {Latham}, \& {Ennico}}]{Deming2009}
{Deming}, D., {et~al.} 2009, \pasp, 121, 952

\bibitem[{{Dotter} {et~al.}(2008){Dotter}, {Chaboyer}, {Jevremovi{\'c}},
  {Kostov}, {Baron}, \& {Ferguson}}]{Dotter2008}
{Dotter}, A., {Chaboyer}, B., {Jevremovi{\'c}}, D., {Kostov}, V., {Baron}, E.,
  \& {Ferguson}, J.~W. 2008, \apjs, 178, 89

\bibitem[{{Dressing} \& {Charbonneau}(2013)}]{Dressing2013}
{Dressing}, C.~D., \& {Charbonneau}, D. 2013, ArXiv e-prints 1302.1647

\bibitem[{{Drimmel} \& {Spergel}(2001)}]{Drimmel2001}
{Drimmel}, R., \& {Spergel}, D.~N. 2001, \apj, 556, 181

\bibitem[{{Everett} {et~al.}(2012){Everett}, {Howell}, \&
  {Kinemuchi}}]{Everett2012}
{Everett}, M.~E., {Howell}, S.~B., \& {Kinemuchi}, K. 2012, \pasp, 124, 316

\bibitem[{{Feiden} \& {Chaboyer}(2012)}]{Feiden2012}
{Feiden}, G.~A., \& {Chaboyer}, B. 2012, \apj, 757, 42

\bibitem[{{Feiden} {et~al.}(2011){Feiden}, {Chaboyer}, \&
  {Dotter}}]{Feiden2011}
{Feiden}, G.~A., {Chaboyer}, B., \& {Dotter}, A. 2011, \apjl, 740, L25

\bibitem[{Fortney {et~al.}(2010)Fortney, Baraffe, \& Militzer}]{Fortney2010}
Fortney, J.~J., Baraffe, I., \& Militzer, B. 2010, in Exoplanets (University of
  Arizon Press), 397

\bibitem[{{Fressin} {et~al.}(2013){Fressin}, {Torres}, {Charbonneau}, {Bryson},
  {Christiansen}, {Dressing}, {Jenkins}, {Walkowicz}, \&
  {Batalha}}]{Fressin2013}
{Fressin}, F., {et~al.} 2013, \apj, 766, 81

\bibitem[{{Gaidos} {et~al.}(2005){Gaidos}, {Deschenes}, {Dundon}, {Fagan},
  {Menviel-Hessler}, {Moskovitz}, \& {Workman}}]{Gaidos2005}
{Gaidos}, E., {Deschenes}, B., {Dundon}, L., {Fagan}, K., {Menviel-Hessler},
  L., {Moskovitz}, N., \& {Workman}, M. 2005, Astrobiology, 5, 100

\bibitem[{{Gaidos} \& {Mann}(2013)}]{Gaidos2013}
{Gaidos}, E., \& {Mann}, A.~W. 2013, \apj, 762, 41

\bibitem[{Gaidos(2000)}]{Gaidos2000}
Gaidos, E.~J. 2000, Icarus, 145, 637

\bibitem[{{Girardi} {et~al.}(2005){Girardi}, {Groenewegen}, {Hatziminaoglou},
  \& {da Costa}}]{Girardi2005}
{Girardi}, L., {Groenewegen}, M.~A.~T., {Hatziminaoglou}, E., \& {da Costa}, L.
  2005, \aap, 436, 895

\bibitem[{{Girardi} {et~al.}(2012){Girardi}, {Barbieri}, {Groenewegen},
  {Marigo}, {Bressan}, {Rocha-Pinto}, {Santiago}, {Camargo}, \& {da
  Costa}}]{Girardi2012}
{Girardi}, L., {et~al.} 2012, {TRILEGAL, a TRIdimensional modeL of thE GALaxy:
  Status and Future}, ed. A.~{Miglio}, J.~{Montalb{\'a}n}, \& A.~{Noels}, 165

\bibitem[{{Greiss} {et~al.}(2012){Greiss}, {Steeghs}, {G{\"a}nsicke},
  {Mart{\'{\i}}n}, {Groot}, {Irwin}, {Gonz{\'a}lez-Solares}, {Greimel},
  {Knigge}, {{\O}stensen}, {Verbeek}, {Drew}, {Drake}, {Jonker}, {Ripepi},
  {Scaringi}, {Southworth}, {Still}, {Wright}, {Farnhill}, {van Haaften}, \&
  {Shah}}]{Greiss2012}
{Greiss}, S., {et~al.} 2012, \aj, 144, 24

\bibitem[{{Hauschildt} {et~al.}(1999){Hauschildt}, {Allard}, \&
  {Baron}}]{Hauschildt1999}
{Hauschildt}, P.~H., {Allard}, F., \& {Baron}, E. 1999, \apj, 512, 377

\bibitem[{{Howard} {et~al.}(2012){Howard}, {Marcy}, {Bryson}, {Jenkins},
  {Rowe}, {Batalha}, {Borucki}, {Koch}, {Dunham}, {Gautier}, {Van Cleve},
  {Cochran}, {Latham}, {Lissauer}, {Torres}, {Brown}, {Gilliland}, {Buchhave},
  {Caldwell}, {Christensen-Dalsgaard}, {Ciardi}, {Fressin}, {Haas}, {Howell},
  {Kjeldsen}, {Seager}, {Rogers}, {Sasselov}, {Steffen}, {Basri},
  {Charbonneau}, {Christiansen}, {Clarke}, {Dupree}, {Fabrycky}, {Fischer},
  {Ford}, {Fortney}, {Tarter}, {Girouard}, {Holman}, {Johnson}, {Klaus},
  {Machalek}, {Moorhead}, {Morehead}, {Ragozzine}, {Tenenbaum}, {Twicken},
  {Quinn}, {Isaacson}, {Shporer}, {Lucas}, {Walkowicz}, {Welsh}, {Boss},
  {Devore}, {Gould}, {Smith}, {Morris}, {Prsa}, {Morton}, {Still}, {Thompson},
  {Mullally}, {Endl}, \& {MacQueen}}]{Howard2012}
{Howard}, A.~W., {et~al.} 2012, \apjs, 201, 15

\bibitem[{Ishiwatari {et~al.}(2007)Ishiwatari, Nakajima, Takehiro, \&
  Hayashi}]{Ishiwatari2007}
Ishiwatari, M., Nakajima, K., Takehiro, S., \& Hayashi, Y.-Y. 2007, Journal of
  Geophysical Research, 112, 1

\bibitem[{{Juri{\'c}} \& {Tremaine}(2008)}]{Juric2008}
{Juri{\'c}}, M., \& {Tremaine}, S. 2008, \apj, 686, 603

\bibitem[{{Kaltenegger}(2010)}]{Kaltenegger2010}
{Kaltenegger}, L. 2010, \apjl, 712, L125

\bibitem[{{Kaltenegger} \& {Sasselov}(2011)}]{Kaltenegger2011}
{Kaltenegger}, L., \& {Sasselov}, D. 2011, \apjl, 736, L25

\bibitem[{Kasting {et~al.}(1993)Kasting, Whitmire, \& Reynolds}]{Kasting1993}
Kasting, J.~F., Whitmire, D.~P., \& Reynolds, R.~T. 1993, Icarus, 101, 108

\bibitem[{{Kipping} {et~al.}(2009){Kipping}, {Fossey}, \&
  {Campanella}}]{Kipping2009}
{Kipping}, D.~M., {Fossey}, S.~J., \& {Campanella}, G. 2009, \mnras, 400, 398

\bibitem[{{Kite} {et~al.}(2009){Kite}, {Manga}, \& {Gaidos}}]{Kite2009}
{Kite}, E.~S., {Manga}, M., \& {Gaidos}, E. 2009, \apj, 700, 1732

\bibitem[{{Kopparapu}(2013)}]{Kopparapu2013b}
{Kopparapu}, R.~K. 2013, ArXiv e-prints 1303.2649

\bibitem[{{Kopparapu} {et~al.}(2013){Kopparapu}, {Ramirez}, {Kasting}, {Eymet},
  {Robinson}, {Mahadevan}, {Terrien}, {Domagal-Goldman}, {Meadows}, \&
  {Deshpande}}]{Kopparapu2013}
{Kopparapu}, R.~K., {et~al.} 2013, \apj, 765, 131

\bibitem[{Kroupa(2002)}]{Kroupa2002}
Kroupa, P. 2002, in Modes of Star Formation and the Origin of Field
  Populations. ASP Conference Series Vol. 285, ed. E.~K. Grebel \& W.~Brandner
  (ASP)

\bibitem[{{Lenz} {et~al.}(1998){Lenz}, {Newberg}, {Rosner}, {Richards}, \&
  {Stoughton}}]{Lenz1998}
{Lenz}, D.~D., {Newberg}, J., {Rosner}, R., {Richards}, G.~T., \& {Stoughton},
  C. 1998, \apjs, 119, 121

\bibitem[{{L{\'e}pine} {et~al.}(2013){L{\'e}pine}, {Hilton}, {Mann}, {Wilde},
  {Rojas-Ayala}, {Cruz}, \& {Gaidos}}]{Lepine2013}
{L{\'e}pine}, S., {Hilton}, E.~J., {Mann}, A.~W., {Wilde}, M., {Rojas-Ayala},
  B., {Cruz}, K.~L., \& {Gaidos}, E. 2013, \aj, 145, 102

\bibitem[{{Mann} {et~al.}(2012){Mann}, {Gaidos}, {L{\'e}pine}, \&
  {Hilton}}]{Mann2012}
{Mann}, A.~W., {Gaidos}, E., {L{\'e}pine}, S., \& {Hilton}, E.~J. 2012, \apj,
  753, 90

\bibitem[{{Moorhead} {et~al.}(2011){Moorhead}, {Ford}, {Morehead}, {Rowe},
  {Borucki}, {Batalha}, {Bryson}, {Caldwell}, {Fabrycky}, {Gautier}, {Koch},
  {Holman}, {Jenkins}, {Li}, {Lissauer}, {Lucas}, {Marcy}, {Quinn}, {Quintana},
  {Ragozzine}, {Shporer}, {Still}, \& {Torres}}]{Moorhead2011a}
{Moorhead}, A.~V., {et~al.} 2011, \apjs, 197, 1

\bibitem[{{Muirhead} {et~al.}(2012){Muirhead}, {Hamren}, {Schlawin},
  {Rojas-Ayala}, {Covey}, \& {Lloyd}}]{Muirhead2012}
{Muirhead}, P.~S., {Hamren}, K., {Schlawin}, E., {Rojas-Ayala}, B., {Covey},
  K.~R., \& {Lloyd}, J.~P. 2012, \apjl, 750, L37

\bibitem[{{Pierrehumbert} \& {Gaidos}(2011)}]{Pierrehumbert2011ApJ734}
{Pierrehumbert}, R., \& {Gaidos}, E. 2011, \apjl, 734, L13

\bibitem[{Pinsonneault {et~al.}(2012)Pinsonneault, An, Molenda-Å»akowicz,
  Chaplan, Metcalfe, \& Bruntt}]{Pinsonneault2012}
Pinsonneault, M., An, D., Molenda-Å»akowicz, J., Chaplan, W.~J., Metcalfe,
  T.~S., \& Bruntt, H. 2012, \apjs, 199, 30

\bibitem[{{Plavchan} {et~al.}(2012){Plavchan}, {Bilinski}, \&
  {Currie}}]{Plavchan2012}
{Plavchan}, P., {Bilinski}, C., \& {Currie}, T. 2012, ArXiv e-prints 1203.1887

\bibitem[{{Reynolds} {et~al.}(1987){Reynolds}, {McKay}, \&
  {Kasting}}]{Reynolds1987}
{Reynolds}, R.~T., {McKay}, C.~P., \& {Kasting}, J.~F. 1987, Adv. Space Res.,
  7, 125

\bibitem[{Schlegel {et~al.}(1998)Schlegel, Finkbeiner, \& Davis}]{Schlegel1998}
Schlegel, D., Finkbeiner, D., \& Davis, M. 1998, \apj, 500, 535

\bibitem[{{Selsis} {et~al.}(2007){Selsis}, {Kasting}, {Levrard}, {Paillet},
  {Ribas}, \& {Delfosse}}]{Selsis2007a}
{Selsis}, F., {Kasting}, J.~F., {Levrard}, B., {Paillet}, J., {Ribas}, I., \&
  {Delfosse}, X. 2007, \aap, 476, 1373

\bibitem[{{Shen} \& {Turner}(2008)}]{Shen2008}
{Shen}, Y., \& {Turner}, E.~L. 2008, \apj, 685, 553

\bibitem[{{Spiegel} {et~al.}(2008){Spiegel}, {Menou}, \&
  {Scharf}}]{Spiegel2008}
{Spiegel}, D.~S., {Menou}, K., \& {Scharf}, C.~A. 2008, \apj, 681, 1609

\bibitem[{Steffen {et~al.}(2012)Steffen, Ford, Rowe, Fabrycky, Holman, Welsh,
  Batalha, Borucki, Bryson, Caldwell, Ciardi, Jenkins, Kjeldsen, Koch,
  Pr\v{s}a, Sanderfer, Seader, \& Twicken}]{Steffen2012}
Steffen, J.~H., {et~al.} 2012, \apj, 756, 186

\bibitem[{{Traub}(2012)}]{Traub2012}
{Traub}, W.~A. 2012, \apj, 745, 20

\bibitem[{Valenti \& Piskunov(1996)}]{Valenti1996}
Valenti, J., \& Piskunov, N. 1996, \aaps, 118, 595

\bibitem[{Vanhollebeke {et~al.}(2009)Vanhollebeke, Groenewegen, \&
  Girardi}]{Vanhollebeke2009}
Vanhollebeke, E., Groenewegen, M. a.~T., \& Girardi, L. 2009, \aap, 498, 95

\bibitem[{{Vickers} {et~al.}(2012){Vickers}, {Grebel}, \&
  {Huxor}}]{Vickers2012}
{Vickers}, J.~J., {Grebel}, E.~K., \& {Huxor}, A.~P. 2012, \aj, 143, 86

\bibitem[{{von Paris} {et~al.}(2013){von Paris}, {Grenfell}, {Hedelt}, {Rauer},
  {Selsis}, \& {Stracke}}]{vonParis2013}
{von Paris}, P., {Grenfell}, J.~L., {Hedelt}, P., {Rauer}, H., {Selsis}, F., \&
  {Stracke}, B. 2013, \aap, 549, A94

\bibitem[{{Wang} \& {Ford}(2011)}]{Wang2011}
{Wang}, J., \& {Ford}, E.~B. 2011, \mnras, 418, 1822

\bibitem[{Williams \& Pollard(2003)}]{Williams2003}
Williams, D.~M., \& Pollard, D. 2003, International Journal of Astrobiology, 2,
  1

\end{thebibliography}

\appendix

\section{Derivation of a uniform probability distribution for extinction}

If the probability distribution of stars with distance $x$ along the
line of sight is $f(x)$ and the density of dust is $g(x)$, then the
total column density of dust along the line of sight to a particular
star is
\begin{equation}
\label{eqn.extinction}
A = A_0  \int_{0}^{x}dx' g(x'),
\end{equation}
where $A_0$ is a constant factor.  The probability of extinction to
any randomly selected star falling between $A$ and $A+dA$ is
\begin{equation}
p(A)dA = f(x)\frac{dx}{dA}dA.
\end{equation}
However, from Eqn. \ref{eqn.extinction}, $dx/dA$ is simply $g(x)^{-1}$
and if $f(x)$ and $g(x)$ are identically distributed with $x$, then
$p(A)$ is a constant, i.e. uniformly distributed over the range of
allowed values.

\clearpage

\begin{deluxetable}{rr|rrrrrrrr|rrrrrr|l}
\tabletypesize{\scriptsize}
\rotate
\tablecaption{Candidate Planets in the Habitable Zones of \emph{Kepler} Stars \label{tab.candidates}}
\tablewidth{0pt}
\tablehead{\colhead{} & \colhead{}  & \multicolumn{8}{c}{Planet Parameters} & \multicolumn{6}{c}{Stellar Parameters} & \colhead{} \\
\colhead{KOI} & \colhead{KIC} & \colhead{$p_{HZ}$} & \colhead{Period} & \multicolumn{3}{c}{Irradiance ($I_{\oplus}$)} & \multicolumn{3}{c}{Radius (\rearth)} & \colhead{$T_*$} & \colhead{log $g$} & \colhead{[Fe/H]} & \colhead{$L_*$} & \colhead{$M_*$} & \colhead{Age} & \colhead{Comment$^b$}\\
\colhead{} & \colhead{} & \colhead{} & \colhead{(days)} & \colhead{MP$^a$} & \colhead{LL$^{a}$} & \colhead{UL$^{a}$} & \colhead{MP$^a$} & \colhead{LL$^{a}$} & \colhead{UL$^{a}$} & \colhead{(K)} &\colhead{}  & \colhead{} & \colhead{($L_{\odot}$)} & \colhead{($M_{\odot}$)} & \colhead{(Gyr)} & \colhead{} }
\startdata
  87.01 & 10593626 & 0.64 &  289.9 &  1.29 &  0.99 &  7.23 &  2.32 &  2.00 &   5.91 & 5735 & 4.43 & -0.2 & 0.90 & 0.93 &  7.8 & Bo11,KS11,Bu12,\emph{Kepler}-22b  \\
 250.04 &  9757613 & 1.00 &   46.8 &  1.33 &  0.99 &  1.53 &  2.16 &  1.91 &   2.22 & 3969 & 4.75 & -0.3 & 0.05 & 0.51 &  8.6 & Ma13,DC13,\emph{Kepler}-26    \\
 351.01 & 11442793 & 0.93 &  331.6 &  1.76 &  1.27 &  2.39 & 10.44 &  9.38 &  12.95 & 6244 & 4.37 & -0.4 & 1.48 & 0.94 &  5.7 & Bo11,KS11               \\
 401.02 &  3217264 & 0.95 &  160.0 &  2.02 &  1.33 &  2.11 &  3.74 &  3.18 &   3.59 & 5528 & 4.52 & -0.2 & 0.62 & 0.89 &  5.5 & Bo11,KS11               \\
 433.02 & 10937029 & 0.99 &  328.2 &  0.82 &  0.51 &  0.88 & 11.46 &  9.20 &  10.65 & 5551 & 4.52 &  0.1 & 0.71 & 1.00 &  1.5 & Bo11,KS11               \\
 463.01 &  8845205 & 0.93 &   18.5 &  1.64 &  1.05 &  2.18 &  1.82 &  1.34 &   2.17 & 3542 & 4.95 & -0.5 & 0.02 & 0.34 &  1.2 & Mu12,Ma13,DC13               \\
 465.01 &  8891318 & 0.55 &  349.9 &  1.67 &  1.03 &  7.05 &  3.63 &  2.89 &   7.77 & 6237 & 4.40 & -0.2 & 1.73 & 1.15 &  1.4 & KS11                    \\
 518.03 &  8017703 & 1.00 &  247.4 &  0.55 &  0.35 &  0.58 &  2.66 &  2.40 &   2.75 & 5045 & 4.64 & -0.5 & 0.26 & 0.69 &  6.4 &                          \\
 622.01 & 12417486 & 0.74 &  155.0 &  1.59 &  1.06 & 22.63 &  5.76 &  4.94 &  26.68 & 5300 & 4.55 & -0.3 & 0.44 & 0.81 &  7.5 & Bo11,KS11               \\
 682.01 &  7619236 & 0.94 &  562.1 &  0.52 &  0.33 &  2.42 &  7.40 &  6.16 &  17.47 & 5918 & 4.51 & -0.3 & 0.92 & 0.99 &  1.4 & KS11,Bu12               \\
 701.03 &  9002278 & 1.00 &  122.4 &  1.20 &  1.02 &  1.50 &  1.79 &  1.78 &   1.98 & 4994 & 4.68 & -0.5 & 0.22 & 0.68 &  1.7 & Bo11,KS11,Bu12          \\
 812.03 &  4139816 & 0.73 &   46.2 &  1.62 &  1.33 &  2.33 &  2.23 &  2.01 &   2.32 & 4029 & 4.72 & -0.4 & 0.07 & 0.57 &  1.5 & Bo11,KS11,Mu12,Ma13     \\
 854.01 &  6435936 & 1.00 &   56.1 &  0.68 &  0.46 &  0.88 &  2.08 &  1.71 &   2.34 & 3661 & 4.80 &  0.0 & 0.03 & 0.49 &  2.1 & Bo11,KS11,Mu12,Ma13,DC13 \\
 881.02 &  7373451 & 0.99 &  226.9 &  0.79 &  0.59 &  1.07 &  3.96 &  3.82 &   4.45 & 5334 & 4.64 & -0.4 & 0.35 & 0.76 &  2.0 & KS11                    \\
 902.01 &  8018547 & 1.00 &   83.9 &  1.46 &  1.10 &  1.75 &  6.55 &  5.56 &   7.04 & 4471 & 4.63 & -0.1 & 0.16 & 0.71 &  5.8 & Bo11,KS11,Mu12          \\
1209.01 &  3534076 & 0.80 &  272.1 &  0.97 &  0.56 &  8.47 &  6.54 &  4.95 &  22.53 & 5587 & 4.54 & -0.4 & 0.59 & 0.84 &  5.5 & Ba13                    \\
1268.01 &  8813698 & 0.51 &  268.9 &  1.83 &  1.15 &  6.16 & 10.24 &  8.75 &  20.25 & 6199 & 4.48 & -0.2 & 1.21 & 0.99 &  2.1 & KS11                    \\
1298.02 & 10604335 & 1.00 &   92.7 &  1.01 &  0.58 &  1.14 &  2.26 &  1.79 &   2.11 & 4337 & 4.67 & -0.3 & 0.13 & 0.68 &  2.0 & Ma13                    \\
1356.01 &  7363829 & 0.89 &  384.0 &  1.10 &  0.63 &  3.08 &  8.97 &  7.16 &  16.36 & 5893 & 4.40 & -0.3 & 1.16 & 0.98 &  5.5 &                          \\
1361.01 &  6960913 & 1.00 &   59.9 &  1.11 &  0.93 &  1.34 &  2.12 &  1.91 &   2.30 & 4070 & 4.74 & -0.3 & 0.07 & 0.54 &  1.9 & Bo11,KS11,Mu12,Ma13     \\
1375.01 &  6766634 & 0.75 &  321.2 &  1.20 &  0.78 &  5.66 &  4.73 &  4.09 &  11.76 & 5989 & 4.47 & -0.5 & 0.92 & 0.86 &  6.0 & Bo11,KS11               \\
1422.02 & 11497958 & 0.98 &   19.9 &  1.50 &  1.24 &  1.58 &  1.39 &  1.29 &   1.39 & 3545 & 4.94 & -0.3 & 0.01 & 0.33 &  8.5 & Mu12,Ma13,DC13          \\
1429.01 & 11030711 & 0.81 &  205.9 &  1.87 &  1.18 &  4.64 &  4.89 &  4.04 &   8.05 & 5719 & 4.47 & -0.3 & 0.82 & 0.91 &  6.0 & Bo11,KS11               \\
1430.03 & 11176127 & 0.98 &   77.5 &  1.69 &  0.87 &  1.93 &  2.81 &  2.14 &   2.62 & 4546 & 4.64 & -0.2 & 0.18 & 0.74 &  1.5 & Ba13                    \\
1431.01 & 11075279 & 0.93 &  345.2 &  0.66 &  0.56 &  2.91 &  6.07 &  5.74 &  14.08 & 5587 & 4.57 & -0.3 & 0.53 & 0.81 &  5.1 & Ba13                    \\
1466.01 &  9512981 & 0.98 &  281.6 &  0.38 &  0.36 &  0.56 & 11.11 & 10.70 &  12.84 & 4763 & 4.63 & -0.2 & 0.23 & 0.77 &  1.5 & Ba13                    \\
1477.01 &  7811397 & 0.93 &  339.1 &  0.46 &  0.40 &  3.41 &  8.64 &  8.20 &  26.53 & 5275 & 4.61 & -0.4 & 0.34 & 0.73 &  6.2 & KS11                    \\
1527.01 &  7768451 & 0.62 &  192.7 &  1.82 &  1.14 & 11.15 &  3.15 &  2.61 &   8.56 & 5733 & 4.53 & -0.2 & 0.75 & 0.96 &  2.0 & Bo11,KS11               \\
1574.02 & 10028792 & 0.99 &  574.0 &  1.09 &  0.70 &  1.67 &  4.99 &  3.77 &   5.28 & 5997 & 4.21 & -0.2 & 2.05 & 1.05 &  6.7 &                          \\
1582.01 &  4918309 & 0.83 &  186.4 &  1.46 &  0.99 & 10.28 &  5.54 &  4.80 &  16.69 & 5571 & 4.58 & -0.4 & 0.54 & 0.87 &  2.0 & Bo11,KS11               \\
1596.02 & 10027323 & 1.00 &  105.4 &  1.28 &  1.21 &  1.69 &  2.68 &  2.39 &   2.59 & 4636 & 4.63 &  0.2 & 0.20 & 0.76 &  1.5 & Bo11,KS11,Mu12          \\
1686.01 &  6149553 & 1.00 &   56.9 &  0.59 &  0.28 &  0.55 &  1.20 &  0.80 &   1.18 & 3597 & 4.83 & -0.0 & 0.03 & 0.46 &  1.4 & Ba13,Ma13,DC13              \\
1739.01 &  7199906 & 0.79 &  220.7 &  1.59 &  0.97 &  7.16 &  1.90 &  1.56 &   4.43 & 5851 & 4.54 & -0.4 & 0.77 & 0.93 &  2.0 & Ba13                    \\
1871.01 &  9758089 & 1.00 &   92.7 &  1.24 &  1.20 &  1.65 &  2.34 &  2.26 &   2.68 & 4534 & 4.66 & -0.3 & 0.16 & 0.70 &  1.5 & Ba13                    \\
1876.01 & 11622600 & 1.00 &   82.5 &  1.28 &  0.76 &  1.53 &  3.04 &  2.25 &   3.26 & 4392 & 4.66 & -0.2 & 0.14 & 0.71 &  1.9 & Ba13                    \\
1879.01 &  8367644 & 0.84 &   22.1 &  1.11 &  1.03 &  2.64 &  1.69 &  1.58 &   2.81 & 3551 & 5.00 & -0.2 & 0.01 & 0.30 &  2.0 & Ma13,DC13                    \\
1902.01 &  5809954 & 0.69 &  137.9 &  0.24 &  0.11 &  0.28 &  2.18 &  1.42 &   2.39 & 3760 & 4.78 & -0.3 & 0.04 & 0.50 &  1.2 & Ba13,Ma13               \\
1986.01 &  8257205 & 0.61 &  148.5 &  1.62 &  1.23 & 16.44 &  3.15 &  2.95 &  11.29 & 5460 & 4.62 & -0.3 & 0.42 & 0.80 &  1.7 &                          \\
1989.01 & 10779233 & 0.65 &  201.1 &  1.80 &  1.38 &  7.21 &  2.19 &  2.04 &   4.93 & 5799 & 4.50 & -0.5 & 0.69 & 0.79 &  8.5 &                          \\
2102.01 &  7008211 & 0.74 &  187.7 &  1.20 &  0.64 & 19.41 &  3.41 &  2.51 &  16.53 & 5307 & 4.56 & -0.4 & 0.42 & 0.78 &  8.0 & Ba13                    \\
2124.01 & 11462341 & 0.64 &   42.3 &  1.74 &  1.54 &  2.71 &  1.06 &  0.98 &   1.31 & 4069 & 4.75 & -0.5 & 0.06 & 0.53 &  2.0 & Ba13,Ma13               \\
2410.01 &  8676038 & 0.58 &  186.7 &  2.08 &  1.59 &  3.36 &  2.05 &  1.89 &   2.74 & 5801 & 4.48 & -0.4 & 0.74 & 0.81 &  8.5 &                          \\
2418.01 & 10027247 & 0.99 &   86.8 &  0.36 &  0.22 &  0.49 &  1.32 &  0.96 &   1.56 & 3739 & 4.84 & -0.1 & 0.03 & 0.46 &  1.9 & Ba13,Ma13,DC13               \\
2469.01 &  6149910 & 0.95 &  131.2 &  0.95 &  0.95 &  1.99 &  1.95 &  2.01 &   2.71 & 4693 & 4.69 & -0.3 & 0.19 & 0.67 &  1.2 & Ba13                    \\
2474.01 &  8240617 & 0.76 &  176.8 &  1.73 &  1.05 & 15.42 &  1.88 &  1.52 &   6.57 & 5589 & 4.54 & -0.4 & 0.59 & 0.84 &  5.5 & Ba13                    \\
2626.01 & 11768142 & 1.00 &   38.1 &  0.84 &  0.51 &  1.06 &  1.26 &  0.92 &   1.43 & 3561 & 4.86 & -0.0 & 0.02 & 0.43 &  1.1 & Ba13,Ma13,DC13          \\
2650.01 &  8890150 & 0.89 &   35.0 &  1.63 &  1.17 &  2.12 &  1.27 &  1.07 &   1.44 & 3855 & 4.78 & -0.1 & 0.05 & 0.51 &  2.0 & Ba13,Ma13,DC13               \\
2681.01 &  6878240 & 0.84 &  135.5 &  1.23 &  1.08 &  2.53 &  4.99 &  4.82 &   6.75 & 5105 & 4.66 & -0.4 & 0.26 & 0.72 &  2.0 &                          \\
2686.01 &  7826659 & 0.96 &  211.0 &  0.48 &  0.46 &  0.62 &  3.28 &  3.15 &   3.62 & 4631 & 4.64 & -0.2 & 0.19 & 0.75 &  1.5 &                          \\
2689.01 & 10265602 & 0.77 &  165.3 &  1.95 &  1.14 & 14.51 &  5.75 &  4.32 &  17.97 & 5593 & 4.53 & -0.4 & 0.60 & 0.84 &  6.1 &                          \\
2691.01 &  4552729 & 0.96 &   97.5 &  1.56 &  1.29 &  1.82 &  3.61 &  3.34 &   3.87 & 4736 & 4.63 & -0.1 & 0.23 & 0.79 &  1.4 &                          \\
2703.01 &  5871985 & 1.00 &  213.3 &  0.40 &  0.37 &  0.48 &  3.35 &  3.16 &   3.56 & 4476 & 4.65 & -0.2 & 0.16 & 0.73 &  1.5 &                          \\
2757.01 &  6432345 & 0.85 &  234.6 &  1.35 &  0.88 &  5.78 &  2.83 &  2.35 &   6.40 & 5735 & 4.54 & -0.3 & 0.71 & 0.93 &  2.1 &                          \\
2762.01 &  8210018 & 0.99 &  133.0 &  0.82 &  0.72 &  1.01 &  2.54 &  2.35 &   2.72 & 4525 & 4.64 & -0.1 & 0.18 & 0.75 &  2.1 &                          \\
2770.01 & 10917043 & 1.00 &  205.4 &  0.39 &  0.32 &  0.47 &  2.51 &  2.22 &   2.71 & 4401 & 4.66 & -0.2 & 0.14 & 0.71 &  2.0 & Ba13                    \\
2834.01 &  5609593 & 0.90 &  136.2 &  0.91 &  0.75 & 21.45 &  2.67 &  2.25 &  13.59 & 4651 & 4.63 & -0.1 & 0.21 & 0.78 &  1.1 &                          \\
2882.01 &  5642620 & 0.55 &   75.9 &  1.49 &  1.40 &  2.58 &  2.28 &  2.11 &   2.73 & 4473 & 4.67 & -0.3 & 0.14 & 0.68 &  1.4 &                          \\
2933.01 & 12416987 & 1.00 &  119.1 &  0.79 &  0.41 &  0.96 &  3.57 &  2.36 &   3.81 & 4411 & 4.66 & -0.2 & 0.14 & 0.71 &  1.0 &                          \\
2992.01 &  8509442 & 1.00 &   82.7 &  0.52 &  0.40 &  0.79 &  2.07 &  1.76 &   2.48 & 3875 & 4.79 & -0.2 & 0.05 & 0.50 &  1.5 &                          \\
3010.01 &  3642335 & 1.00 &   60.9 &  0.76 &  0.60 &  1.05 &  1.37 &  1.19 &   1.58 & 3845 & 4.79 & -0.1 & 0.04 & 0.50 &  2.0 &                          \\
3034.01 &  2973386 & 0.66 &   31.0 &  1.68 &  1.23 &  2.42 &  1.49 &  1.21 &   1.77 & 3825 & 4.82 & -0.2 & 0.04 & 0.48 &  1.5 &                          \\
3086.01 & 10749059 & 0.89 &  174.7 &  1.31 &  1.01 &  9.64 &  2.75 &  2.57 &   8.46 & 5462 & 4.62 & -0.3 & 0.42 & 0.81 &  1.7 &                          \\
\hline
\multicolumn{17}{c}{Other Planets with $R_p < 2$\rearth{} and $p_{HZ} > 0.01$}\\ 
 172.02 &  8692861 & 0.38 &  242.5 &  2.44 &  1.63 &   4.81 &  1.88 &  1.64 &   2.44 & 6140 & 4.39 & -0.2 & 1.38 & 0.97 &  5.1 &                          \\
 775.03 & 11754553 & 0.15 &   36.4 &  2.12 &  1.84 &   2.55 &  1.81 &  1.67 &   1.96 & 4061 & 4.74 & -0.3 & 0.06 & 0.54 &  1.9 & Ma13                    \\
 817.01 &  4725681 & 0.03 &   24.0 &  3.29 &  1.99 &   3.62 &  1.99 &  1.57 &   2.07 & 3900 & 4.73 & -0.0 & 0.06 & 0.57 &  1.0 & Bo11,Ma13,Mu12            \\
1078.03 & 10166274 & 0.47 &   28.5 &  1.67 &  1.57 &   3.04 &  1.88 &  1.81 &   2.31 & 3790 & 4.84 & -0.5 & 0.03 & 0.45 &  1.1 & Ma13                    \\
2179.01 & 10670119 & 0.09 &   14.9 &  3.09 &  1.84 &   4.03 &  1.32 &  0.95 &   1.54 & 3606 & 4.87 & -0.2 & 0.02 & 0.42 &  1.5 & Ma13                    \\
2339.02 &  7033233 & 0.05 &   65.2 &  2.04 &  1.99 &   2.79 &  1.43 &  1.32 &   1.43 & 4551 & 4.66 & -0.3 & 0.16 & 0.71 &  1.4 &                          \\
2373.01 & 10798331 & 0.16 &  147.3 &  2.16 &  1.89 &   7.89 &  1.96 &  1.86 &   3.84 & 5590 & 4.55 & -0.2 & 0.56 & 0.82 &  5.9 &                          \\
2760.01 &  7877978 & 0.22 &   56.6 &  2.47 &  1.35 &   2.92 &  1.92 &  1.30 &   2.03 & 4510 & 4.65 & -0.2 & 0.17 & 0.74 &  1.5 &                          \\
2862.01 &  6679295 & 0.14 &   24.6 &  2.84 &  1.71 &   3.45 &  1.72 &  1.32 &   1.85 & 3823 & 4.74 &  0.0 & 0.05 & 0.55 &  2.0 &                          \\
2931.01 &  8611257 & 0.39 &   99.2 &  2.47 &  1.65 &  24.73 &  1.95 &  1.61 &   7.29 & 5129 & 4.56 & -0.2 & 0.38 & 0.80 &  7.5 &                          \\
\enddata
%% Text for table notes should follow after the \enddata but before
%% the \end{deluxetable}. Make sure there is at least one \tablenotemark
%% in the table for each \tablenotetext.
%\tablecomments{This is a table comment.}
\tablenotetext{a}{MP = most probable value, LL = 95\% lower limit, UL = 95\% upper limit}
\tablenotetext{b}{Reported as HZ candidate in: Bo11 = \citet{Borucki2011ApJ736}, KS12 = \citet{Kaltenegger2011}, Ba13 = \citet{Batalha2013}; DC13 = \citet{Dressing2013}.  Spectroscopy reported in Bu12 = \citet{Buchhave2012}, Mu12 = \citet{Muirhead2012}, Ma13 = Mann et al., in prep.}
\end{deluxetable}

\end{document}